\documentclass[a4paper,onecolumn,11pt,accepted=2022-10-16]{quantumarticle}
\pdfoutput=1

\usepackage{amsmath,amssymb,amsthm,a4wide}
\usepackage{latexsym}
\usepackage{indentfirst}
\usepackage{graphicx}
\usepackage{placeins}
\usepackage{booktabs}
\usepackage{algorithm}
\usepackage{algorithmic}
\usepackage{multirow}
\usepackage{color}

\renewcommand{\d}{\mathrm{d}}

\newcommand{\eps}{\epsilon}

\newcommand{\bbm}{\begin{bmatrix}}
\newcommand{\ebm}{\end{bmatrix}}
\newcommand{\R}{\mathbb{R}}

\newcommand{\D}{\mathbb{D}}

\newcommand{\Z}{\mathbb{Z}}

\newcommand{\pD}{\gamma}

\newcommand{\z}{\xi}

\newcommand{\hg}{\hat{g}}

\newcommand{\dz}{\mathrm{d}z}

\newcommand{\dzoz}{\frac{\dz}{z}}

\renewcommand{\Re}{\mathrm{Re}}
\renewcommand{\Im}{\mathrm{Im}}

\newcommand{\eve}{\text{e}}
\newcommand{\odd}{\text{o}}

\begin{document}

\title{Stable factorization for  phase factors of quantum signal processing}

\author{Lexing Ying}
\affiliation{Department of Mathematics, Stanford University, Stanford, CA 94305, USA}
\orcid{0000-0003-1547-1457}

\maketitle

\begin{abstract}
  This paper proposes a new factorization algorithm for computing the phase factors of quantum signal
  processing. The proposed algorithm avoids root finding of high degree polynomials by using a key
  step of Prony's method and is numerically stable in the double precision arithmetics. Experimental
  results are reported for Hamiltonian simulation, eigenstate filtering, matrix inversion, and
  Fermi-Dirac operator.
\end{abstract}



\section{Introduction}\label{sec:intro}

\subsection{Background}
This paper is concerned with the problem of quantum signal processing. Quantum computing has been
mostly working with unitary operators, since the quantum gates and circuits are unitary. However, in
recent years, we have witnessed great progress in representing non-unitary operators efficiently
with quantum circuits. 

Let $A$ be an $N\times N$ Hermitian matrix with $N = 2^n$ and $\|A\|_2<1$ (after scaling if
  needed). For simplicity, we only consider Hermitian matrices in this paper and refer the readers
  to \cite{gilyen2019quantum,dong2021efficient} for more general cases.  One of the most successful
methods for presenting $A$ on a quantum circuit is the Hermitian block encoding
\[
A \leftrightarrow
\begin{bmatrix}
  A & *\\
  * & *
\end{bmatrix}
\equiv U_A,
\]
where $U_A$ is a Hermitian unitary matrix of size $(2^m\cdot N) \times (2^m\cdot N)$, $A$ is the
top-left corner of $U_A$, and $U_A$ can be implemented using a quantum circuit with $n+m$ input
qubits.

In most of the quantum problems in scientific computing, such as Hamiltonian simulation, filtering,
and quantum linear algebra \cite{childs2018toward,martyn2021grand,dong2021efficient,van2020quantum},
one is often interested the Hermitian matrix $f(A)$ of $A$, where $f(x)$ is a real function defined
on $[-1,1]$ with $\|f\|_\infty < 1$. The block encoding scheme requires $f(A)$ to be represented as
the top-left block of a larger unitary matrix $U_{f(A)}$ implemented by a quantum circuit
\[
f(A) \leftrightarrow
\begin{bmatrix}
  f(A) & *\\
  * & *
\end{bmatrix}
\equiv U_{f(A)}.
\]

\begin{figure}
  \centering
  \begin{tabular}{cc}
    \includegraphics[scale=0.32]{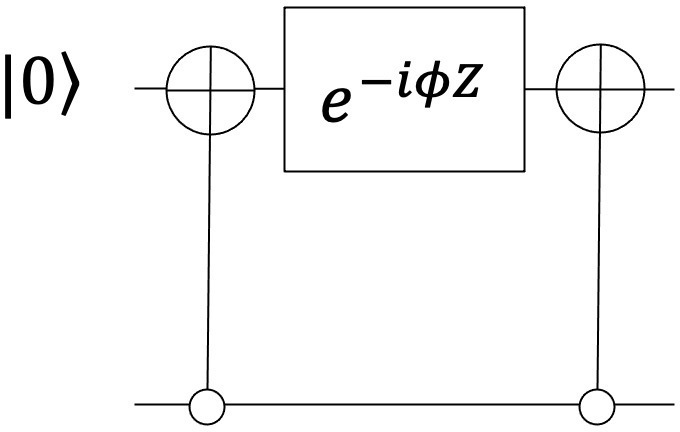}  &     \includegraphics[scale=0.32]{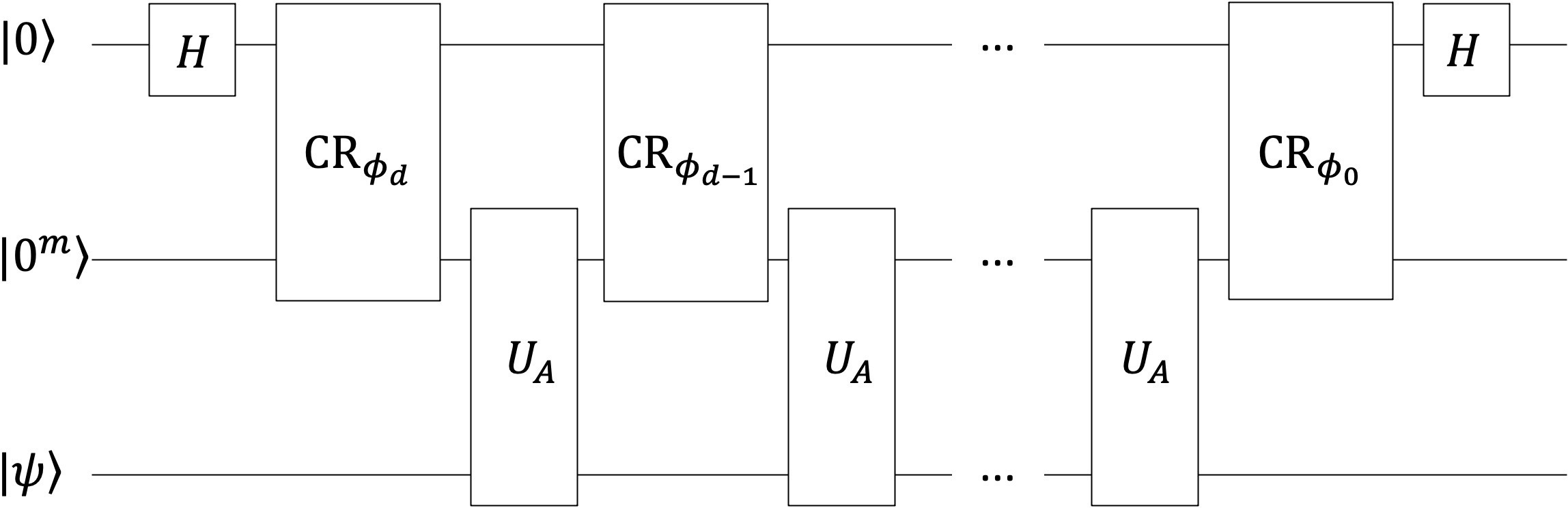} \\
    (a) & (b)
  \end{tabular}
  \caption{
    (a) a controlled rotation circuit $\text{CR}_{\phi}$ with angle $\phi$. (b) quantum
    eigenvalue transformation. Here $H$ is the Hadamard gate, $U_A$ is the block encoding of a
    Hermitian $A$, and $\text{CR}_{\phi_j}$ is with the control angle $\phi_j$. The whole circuit
    implements $U_{a(A)}$ for a real polynomial $a(x)$.
  }
  \label{fig:circuit}
\end{figure}

A key question is whether there is an algorithm that builds the quantum circuit $U_{f(A)}$ from the
circuit $U_A$ by using only the knowledge of the function $f(x)$ but treating $U_A$ as a black box
\[
U_A \equiv
\begin{bmatrix}
  A & *\\
  * & *
\end{bmatrix}
\Rightarrow
U_{f(A)} \equiv 
\begin{bmatrix}
  f(A) & *\\
  * & *
\end{bmatrix}.
\]
This question is answered by the quantum eigenvalue transformation described in
\cite{low2017optimal,gilyen2019quantum}. To simplify the
discussion, we assume that all Hermitian matrices mentioned below satisfy $\|A\|_2< 1$ and all
functions defined on $[-1,1]$ satisfy $\|f\|_\infty < 1$. The quantum eigenvalue transform proceeds
as follows (see \cite{lin2022lecture} for example for details).
\begin{itemize}
\item Split the polynomial $f(x)$ into the even and odd parts $f^\eve(x)$ and $f^\odd(x)$ on
  $x\in[-1,1]$
\item Approximate the even part $f^\eve(x)$ with an even degree polynomial $a^\eve(x)$ and implement
  $a_\eve(A)$ with a circuit shown in Figure \ref{fig:circuit}(b) with appropriate phase factors
  $\phi^\eve_0,\ldots,\phi^\eve_{d_\eve}$. Here $d_\eve$ is the equal to degree of $a^\eve(x)$.
\item Approximate the odd part $f^\odd(x)$ with an odd degree polynomial $a^\odd(x)$ and implement
  $a_\odd(A)$ with a circuit shown in Figure \ref{fig:circuit}(b) with appropriate phase factors
  $\phi^\odd_0,\ldots,\phi^\odd_{d_\odd}$. Here $d_\odd$ is the equal to degree of $a^\odd(x)$.
\item Combine the circuits implementing each component together by linear combination of unitaries
  (LCU) \cite{childs2017quantum}.
\end{itemize}

The key remaining step is how to construct the phase factors $\phi_0,\ldots,\phi_d$ for an even or odd
polynomial $a(x)$ of degree $d$. This is answered by the quantum signal processing theorem
\cite{low2017optimal,gilyen2019quantum,martyn2021grand}: Given a polynomial $a(x)\in\R[x]$ on
$[-1,1]$ of degree $d$, parity $d \mod 2$, and $\|a\|_\infty=\max_{x\in[-1,1]}|a(x)|<1$, there
exists a sequence of phase factors $\Phi=(\phi_0,\ldots,\phi_d) \in [-\pi,\pi]^{d+1}$ such that
$a(x) = \Re(p(x))$, where $p(x)$ is defined via
\begin{equation}
  \label{eq:U}
  U(x,\Phi)=
\begin{pmatrix}
  p(x) & r(x)\\
  r^*(x) & p^*(x)
\end{pmatrix}
= e^{i\phi_0 Z} e^{i\arccos(x) X} e^{i\phi_1 Z} e^{i\arccos(x) X} \cdots e^{i\phi_{d-1} Z} e^{i\arccos(x) i X} e^{i\phi_d Z},
\end{equation}
where 
\[
X = \begin{pmatrix}
  0 & 1\\
  1 & 0
\end{pmatrix},
\quad
Z = \begin{pmatrix}
  1 & 0\\
  0 & -1
\end{pmatrix}
\]
are the Pauli matrices.

In this paper, we use the following notational convention
\begin{equation}
  a(x) = \Re(p(x)), \quad
  c(x) = \Im(p(x)), \quad
  b(x) = \Re(r(x)), \quad
  d(x) = \Im(r(x)).
  \label{eq:abcd}
\end{equation}

It is often convenient to work with variable $t\in[-\pi,\pi]$ and lift these functions to the $t$
space via the transform $x=\cos(t)$, i.e., given $f(x)$ for $x\in[-1,1]$, define
\[
f(t) := f(x=\cos(t))
\]
for $t\in [0,\pi]$ and extend to $t\in[-\pi,0]$ analytically. For example $f(x)=x$ lifts to
$f(t)=\cos(t)$ and $f(x) = \sqrt{1-x^2}$ to $f(t)=\sin(t)$. In the $t$ variable, \eqref{eq:U} can be
written more compactly as
\begin{equation}
  \label{eq:Ut}
  U(t,\Phi)=
\begin{pmatrix}
  p(t) & r(t)\\
  r^*(t) & p^*(t)
\end{pmatrix}
= e^{i\phi_0 Z} e^{i t X} e^{i\phi_1 Z} e^{i t X} \cdots e^{i\phi_{d-1} Z} e^{i t X} e^{i\phi_d Z}.
\end{equation}

We can also use complex variable $z = e^{i t}$ and lift $f(t)$ analytically to a Laurent polynomial
$f(z)$ with
\[
f(z=e^{it})  = f(t) 
\]
on the unit circle. For example $f(t)=\cos(t)$ lifts to $f(z)=\frac{z+z^{-1}}{2}$ and $f(t)=\sin(t)$
to $f(z)=\frac{z-z^{-1}}{2i}$. In the following discussion, we often work with the lifted functions
$a(z)$, $b(z)$, $c(z)$, and $d(z)$ over the complex plane.

\subsection{Previous work}
There are two main approaches for computing the phase factors. The first one
\cite{gilyen2019quantum,haah2019product,chao2020finding} is based on polynomial factorization.
Following the notation of \cite{haah2019product}, this approach starts by choosing a function
$b(\cdot)$ that has the right parity and satisfies $a^2(t)+b^2(t)<1$. Let $\{\z_j\}$ be the set of
$2d$ roots of the Laurent polynomial $1-a^2(z)-b^2(z)$ inside the unit circle. Define
\[
e(z) = z^{-d} \prod_{|\z_j|< 1} (z-\z_j).
\]
By setting $\alpha\equiv\frac{1-a^2(z)-b^2(z)}{e(z) e(1/z)}$, the functions $c(z)$ and $d(z)$ are
then equal to \cite{haah2019product}
\begin{equation}
  c(z) = \left( \sqrt{\alpha}\cdot \frac{e(z)+e(1/z)}{2}  \right),\quad
  d(z) = \left( \sqrt{\alpha}\cdot \frac{e(z)-e(1/z)}{2i} \right).
  \label{eq:cd}
\end{equation}
With $a(z)$, $b(z)$, $c(z)$, and $d(z)$ available, $p(z)=a(z)+ic(z)$ and $r(z)=b(z)+id(z)$ as
defined in \eqref{eq:abcd}. Given $p(z)$ and $r(z)$, the algorithm for extracting the phase factors
$\Phi=(\phi_0,\ldots,\phi_d)$ is quite straightforward (see for example Theorem 3 of
\cite{gilyen2018quantum}). For completeness, it is also included in Section \ref{sec:idea} following
our notation.

Though this approach is direct, the implementation requires finding roots of a high degree Laurent
polynomial, which is often unstable in double precision arithmetics. It was shown that
\cite{haah2019product} that $O(d \log d/\eps)$ classical bits are needed and the algorithms are
often implemented with variable precision. In \cite{chao2020finding}, an algorithm based on the
halving and capitalization techniques is proposed to mitigate the numerical issue and it was able to
scales to more than $3000$ phase factors.

The second approach is based on optimization \cite{dong2021efficient}, i.e., minimizing directly
\[
\min_\Phi \int_{[-1,1]} |\Re (U(x,\Phi)_{11}) - a(x)|^2 \d x,
\]
but with the search space restricted to the symmetric phase factors $\Phi$ (i.e.
$\phi_j=\phi_{d-j}$). Though this minimization problem is highly non-convex,
\cite{dong2021efficient} demonstrates numerically that, starting from the initial guess $\Phi^0 =
(\pi/4,0,\ldots,0,\pi/4)$, a quasi-Newton method is able to find the {\em maximal solution} that
corresponds to $b(t)=0$ in our notation. The numerical results in \cite{dong2021efficient}
demonstrated robust computation of the phase factors up to $10000$ phase factors. A recent study
\cite{wang2021energy} proves that for $\|a\|_\infty \le O(1/d)$ a projected gradient method
converges to the maximal solution.


\subsection{Contribution}

The main contribution of this paper is a new stable algorithm of the factorization approach. It is
based on two observations. First, the factorization approach does not really need the roots
$\{\z_j\}$ since the function $e(z)=z^{-d}\prod_{|\z_j|<1}(z-\z_j)$ only depends on the
characteristic polynomial $\prod_{|\z_j|<1}(z-\z_j)$ of these roots. We show that this
characteristic polynomial can be computed directly via a key component of Prony's method
\cite{prony1795essai,potts2013parameter}, without knowing the roots $\{\z_j\}$. This avoids the
root-finding, which is the main source of instability of the factorization approach. Second, in
order to compute the characteristic polynomial in a robust way, we propose to pick $b(z)$ randomly
with a dominant highest frequency, i.e., in some sense opposite to the symmetric phase factors. This
allows us to compute the characteristic polynomial using the standard numerical linear algebra
routines.

The resulting algorithm is conceptually simple and easy to implement. On the numerical side,
compared with the-state-of-the-art results in \cite{dong2021efficient}, our algorithm achieves
comparable accuracy ($\sim 10^{-12})$ and has the same $O(d^2)$ computational cost. The longest
sequence reported in our experiments scales to over $50000$ phase factors.

The rest of the paper is organized as follows. Section \ref{sec:prony} reviews the Prony's method.
Section \ref{sec:algo} describes the main algorithms. The numerical results are given in Section
\ref{sec:res}.

\section{Review of Prony's method}\label{sec:prony}

Let us explain Prony's method with a simple but key example. Let $(f_k)_{k\in\Z}$ be a sequence of
the form
\[
f(k) = \sum_{j=1}^d e^{i \omega_j k} r_j, 
\]
where $d$ is the number of terms, $\{\omega_j\}$ are the frequencies, and $\{r_j\}$ are the weights.
Assume that $d$, $\{\omega_j\}$, and $\{r_j\}$ are all unknown to us. The computation problem is to
recover $d$, $\{\omega_j\}$ (up to $2\pi$), and $\{r_j\}$ from potentially noisy values of
$(f_k)_{k\in\Z}$.

Prony's method starts by considering the infinite vector $[e^{i \omega_j k}]_{k\in \Z}$ for some
$j$ and the upward shift operator $S$. Applying $S$ to this vector gives
\[
S
\begin{bmatrix}
  \vdots\\
  e^{i \omega_j k}\\
  \vdots
\end{bmatrix}
=
\begin{bmatrix}
  \vdots\\
  e^{i \omega_j (k+1)}\\
  \vdots
\end{bmatrix} \quad\text{i.e.}\quad
(S-e^{i \omega_j})
\begin{bmatrix}
  \vdots\\
  e^{i \omega_j k}\\
  \vdots
\end{bmatrix}
=0.
\]
Taking the product over all $(S-e^{i \omega_j})$ leads to
\[
\prod_{s=1}^d (S-e^{i \omega_s})
\begin{bmatrix}
  \vdots\\
  e^{i \omega_j k}\\
  \vdots
\end{bmatrix}
=0.
\]
Taking a linear combination of the vectors over $j$ with unknown weights $r_j$ gives
\[
\prod_{s=1}^d (S-e^{i \omega_s})
\begin{bmatrix}
  \vdots\\
  \sum_{j=1}^d e^{i \omega_j k} r_j\\
  \vdots
\end{bmatrix}
=0
\Rightarrow
\prod_{s=1}^d (S-e^{i \omega_s})
\begin{bmatrix}
  \vdots\\
  f_k\\
  \vdots
\end{bmatrix}
=0.
\]
Define the polynomial $m(z) \equiv m_0 + \ldots + m_d z^d\equiv\prod_{s=1}^d(z-e^{i\omega_s})$. Then
the last equality becomes
\begin{equation}
m(S) \begin{bmatrix}    \vdots\\    f_k\\    \vdots  \end{bmatrix}\equiv
m_0 \cdot S^0 \begin{bmatrix}    \vdots\\    f_k\\    \vdots  \end{bmatrix} +
\cdots +
m_d \cdot S^d \begin{bmatrix}    \vdots\\    f_k\\    \vdots  \end{bmatrix} = 0
\Rightarrow
\begin{bmatrix}
  \vdots &\vdots &\cdots &\vdots \\    
  f_k & f_{k+1} & \cdots & f_{k+d} \\    
  \vdots &\vdots &\cdots &\vdots 
\end{bmatrix}
\begin{bmatrix}
  m_0\\
  \ldots\\
  m_d
\end{bmatrix}
= 0.
\label{eq:pronyls}
\end{equation}
The final linear system contains a great deal of information.
\begin{itemize}
\item The rank of the matrix in \eqref{eq:pronyls} gives $d$.
\item Any non-zero vector in the null space of the matrix in \eqref{eq:pronyls} gives the
  coefficients of $m_0,\ldots,m_d$ of the polynomial $m(z)$.
\item The roots of $m(z)$ gives $\{e^{i \omega_j} \}$.
\item Solving the least-squares problem
  \[
  \min_{r_j} \sum_k \left|\sum_{j=1}^d e^{i \omega_j k} r_j - f_k \right|^2
  \]
  gives $\{r_j\}$.
\end{itemize}
Though we describe Prony's method using infinite vectors, it is clear now that only $d+1$ rows of
the matrix is needed. Due to the shifting nature of the matrix, only $2d+1$ consecutive values of
$(f_k)$ are required.

The main advantages of the Prony's method are that (1) it is adaptive in the sense that
$\{\omega_j\}$ do not need to fall in any discrete grid, (2) it is conceptually simple, and (3) it
leverages standard numerical routines such as root-finding and null-space computation.  The main
disadvantage is that root-finding can often be unstable when noise is present.

\section{Algorithm}\label{sec:algo}

\subsection{Key components}\label{sec:idea}

We start by choosing the function $b(t)$ to be of the form
\begin{equation}
  b(t) =  b_d \sin(dt) + b_{d-2} \sin((d-2)t) + \ldots.
  \label{eq:b}
\end{equation}
Here the leading coefficient is the most dominant one and the rest of the coefficients
$(b_{d-2},\ldots)$ are chosen randomly. The reason for doing so will be explained below.

Recall that the key function of the factorization approach is $e(z)=z^{-d}\prod_{|\z_j|<1}(z-\z_j)$,
where the second term $\prod_{|\z_j|< 1}(z-\z_j)$ is the characteristic polynomial of the $2d$ roots
$\{\z_j\}$ of $1-a^2(z)-b^2(z)$ inside the unit circle.

\subsubsection{Characteristic polynomial.}  The first idea is that it is possible to compute the
characteristic polynomial directly without first calculating the roots. This avoids the
root-finding, which is the main source of instability of the factorization approach. A simple but
key observation is that these roots are the poles of the reciprocal $g(z)=
\left(1-a^2(z)-b^2(z)\right)^{-1}$ inside the unit disk.


Since $g(z)$ is meromorphic, $g(z)$ takes the form
\[
g(z) = \sum_{\z_j} \frac{w_j}{\z_j-z} + \text{constant},
\]
where the sum is taken over the roots both inside and outside $\D$. Let us consider the integrals
\begin{equation}
  \frac{1}{2\pi i}\int_{\pD} \frac{g(z)}{z^k} \dzoz
  \label{eq:integral}
\end{equation}
for integer values of $k\le -1$, where $\pD$ is the boundary of $\D$ in the counter clockwise
orientation. For a fixed $k\le -1$,
\[
\begin{aligned}
  & \frac{1}{2\pi i}\int_{\pD} \frac{g(z)}{z^k}\dzoz  =
  \frac{1}{2\pi i}\int_{\pD} \left( \sum_{|\z_j|<1} + \sum_{|\z_j|>1}\right) \frac{w_j}{\z_j-z} z^{-(k+1)} \dz \\
  &= \frac{1}{2\pi i} \sum_{|\z_j|<1} w_j \int_{\pD} \frac{1}{\z_j-z} z^{-(k+1)} \dz
  = \frac{1}{2\pi i} \sum_{|\z_j|<1} w_j \z_j^{-(k+1)} \int_{\pD} \frac{1}{\z_j-z} \dz = - \sum_{|\z_j|<1} w_j \z_j^{-(k+1)},
\end{aligned}
\]
where the second equality relies on the analyticity of $\frac{w_j}{\z_j-z}$ in $\D$ for $|\z_j|>1$
and the third equality uses the residue theorem at $\{\z_j\}$. This computation shows that the
integrals $\frac{1}{2\pi i}\int_{\pD} g(z)z^k \dz$ for $k\le -1$ contain important information about
the poles inside $\D$.

The integral $\frac{1}{2\pi i}\int_{\pD} \frac{g(z)}{z^k} \dzoz$ over the unit circle is also
closely related to the Fourier transform of the function $g(t) \equiv g(e^{it})$:
\begin{equation}
  \frac{1}{2\pi i}\int_{\pD} \frac{g(z)}{z^k} \dzoz = \frac{1}{2\pi i}\int_0^{2\pi} g(t) e^{-ik t} i \d t
  = \frac{1}{2\pi} \int_0^{2\pi} g(t) e^{-ikt} \d t = \hg_k.
  \label{eq:Fourier}
\end{equation}

In order to recover the characteristic polynomial $\prod_{|\z_j|< 1} (z-\z_j)$ (the key part of
$e(z)$), we apply Prony's method to the Fourier coefficients. Slightly different from the
description in Section \ref{sec:prony}, we define the semi-infinite (instead of infinite) vector
\[
\hg_-
\equiv
\begin{bmatrix}
  \hg_{-1}\\
  \hg_{-2}\\
  \vdots
\end{bmatrix}
\equiv
\frac{1}{2\pi i}\int_{\pD} g(z) 
\begin{bmatrix}
  z^0\\
  z^1\\
  \vdots
\end{bmatrix}
\d z
\equiv
\begin{bmatrix}
  -\sum_{|\z_j|<1} w_j \z_j^0\\
  -\sum_{|\z_j|<1} w_j \z_j^1\\
  \vdots
\end{bmatrix}
\]
Let $S$ be the shift operator that shifts the semi-infinite vector upward (i.e., dropping the first
element). For any $\z_j$ with $|\z_j|<1$,
\[
S
\begin{bmatrix}
  \z_j^0\\
  \z_j^1\\
  \vdots
\end{bmatrix}
=
\begin{bmatrix}
  \z_j^1\\
  \z_j^2\\
  \vdots
\end{bmatrix},
\quad\text{i.e.,}\quad
(S-\z_j)
\begin{bmatrix}
  \z_j^0\\
  \z_j^1\\
  \vdots
\end{bmatrix}
= 0.
\]
Since the operators $S-\z_j$ all commute, 
\begin{equation}
\prod_{|\z_i|<1} \left(S- \z_i\right)
\begin{bmatrix}
  \z_j^0\\
  \z_j^1\\
  \vdots
\end{bmatrix}
= 0.
\label{eq:prodR}
\end{equation}
Since $\hg_-$ is a linear combination of such semi-infinite vectors with weights $\{-w_j\}$,
\[
\prod_{|\z_i|<1} \left(S- \z_i\right) \hg_- = 0.
\]
Since $b(z)$ is chosen randomly, with probability 1 the roots $\{\z_i\}$ are disjoint. Therefore,
the polynomial $\prod_{|\z_i|<1} \left(z-\z_i\right)$ is of degree $2d$. By denoting it as
\[
m(z) = m_0 z^0 + \cdots + m_{2d} z^{2d}, 
\]
\eqref{eq:prodR} becomes $m_0 (S^0 \hg_-) + \cdots + m_{2d} (S^{2d} \hg_-) = 0$, i.e.,
\begin{equation}
  \begin{bmatrix}
    \hg_{-1} & \hg_{-2} & \cdots & \hg_{-(2d+1)} \\
    \hg_{-2} & \hg_{-3} & \cdots & \hg_{-(2d+2)} \\
    \vdots & \vdots & \ddots & \vdots
  \end{bmatrix}
  \begin{bmatrix}
    m_0\\
    \ldots\\
    m_{2d}
  \end{bmatrix}
  = 0.
  \label{eq:ls}
\end{equation}
At this point, $(m_0,\ldots,m_{2d})$ can be computed as a non-zero vector in the null-space of the
matrix in \eqref{eq:ls}. Once $m(z)$ is obtained, we set $e(z) = z^{-d} m(z)$ as defined. Once
$e(z)$ is ready, the Laurent polynomials $c(z)$, $d(z)$, $p(z)=a(z)+i c(z)$, and $r(z)=b(z)+id(z)$
follow from \eqref{eq:cd}.

\subsubsection{Phase factors from $p(z)$.}\label{sec:frompz}

The construction of the actual phase factors is given as follows,
essentially following Theorem 3 of \cite{gilyen2018quantum} but in terms of the $t$ variable.

For each $n=d$ down to $0$, perform the following two steps
\begin{itemize}
\item In the $t$ variable, $p(t)$ and $r(t)$ are trigonometric polynomials of degree $n$.  Write
  $p(t) = p_n e^{int} + \ldots$ and $r(t) = r_n e^{int} + \ldots$, where $p_n$ and $r_n$ are the
  degree $n$ coefficients. Solve $\phi_n$ from $e^{2i\phi_n} = p_n/r_n$.
\item Transform $p(t)$ and $r(t)$ via
  \begin{equation}
  \begin{pmatrix}
    p(t) & r(t)
  \end{pmatrix}
  \Leftarrow
  \begin{pmatrix}
    p(t) & r(t)
  \end{pmatrix}
  \begin{pmatrix}
    e^{-i\phi_n} & 0\\
    0 & e^{i\phi_n}
  \end{pmatrix}
  \begin{pmatrix}
    \cos(t) & -i\sin(t)\\
    -i\sin(t) & \cos(t)
  \end{pmatrix}.
  \label{eq:pr}
  \end{equation}
  This brings the top coefficients of $p(t)$ and $r(t)$ to zero, hence reducing the degree by one.
\end{itemize}
Within this loop, the switch between the angular function $p(t)$ and the coefficients
$\{p_j\}_{-n\le j \le n}$ can be done with the fast Fourier transform (FFT). Because of the $O(d\log
d)$ complexity of the FFT, the overall cost of this loop is $O(d^2 \log d)$.

\subsubsection{Robust computation of polynomial coefficients}.  The remaining issue is to compute
$(m_0,\ldots,m_{2d})$ in a numerically stable way. This is in fact not always guaranteed. Consider
for example the case that $a(z)$ has negligible coefficients for large frequency. If we set
$b(z)=0$, then $g(z) = (1-a^2(z)-b^2(z))^{-1} = (1-a^2(z))^{-1}$ might lack high frequency
content. This implies that all coefficients $\hg_k$ might be be negligible for large $k$ values. A
direct consequence is that the matrix in \eqref{eq:ls} might have a numerical rank much smaller than
$2d$. In order to resolve this issue, we choose $b(z)$ to have a large leading coefficient as
suggested in \eqref{eq:b}. This is the second contribution of this paper.

\begin{figure}[h!]
  \centering
  \begin{tabular}{cc}
    \includegraphics[scale=0.3]{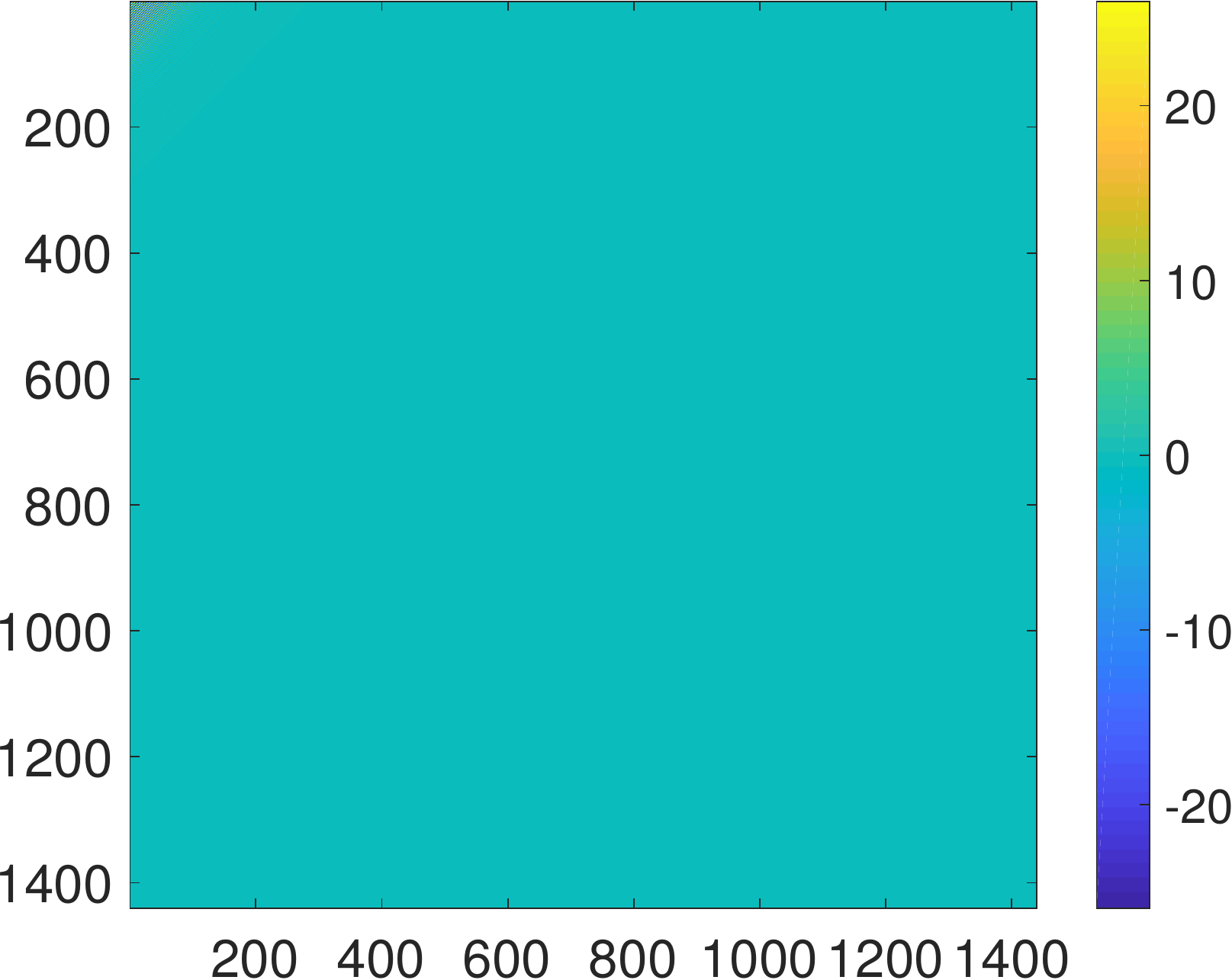} & \includegraphics[scale=0.3]{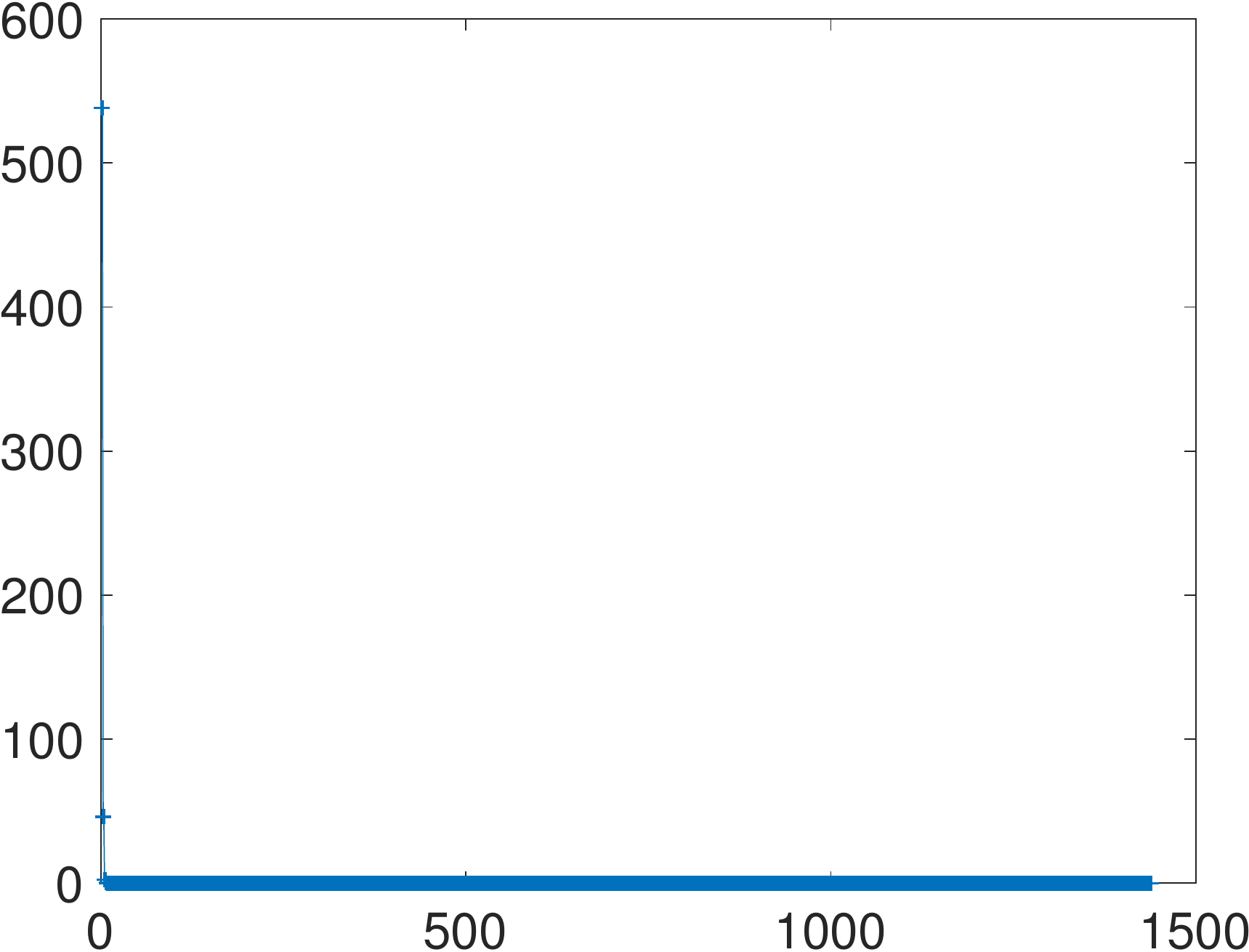}\\
    \includegraphics[scale=0.3]{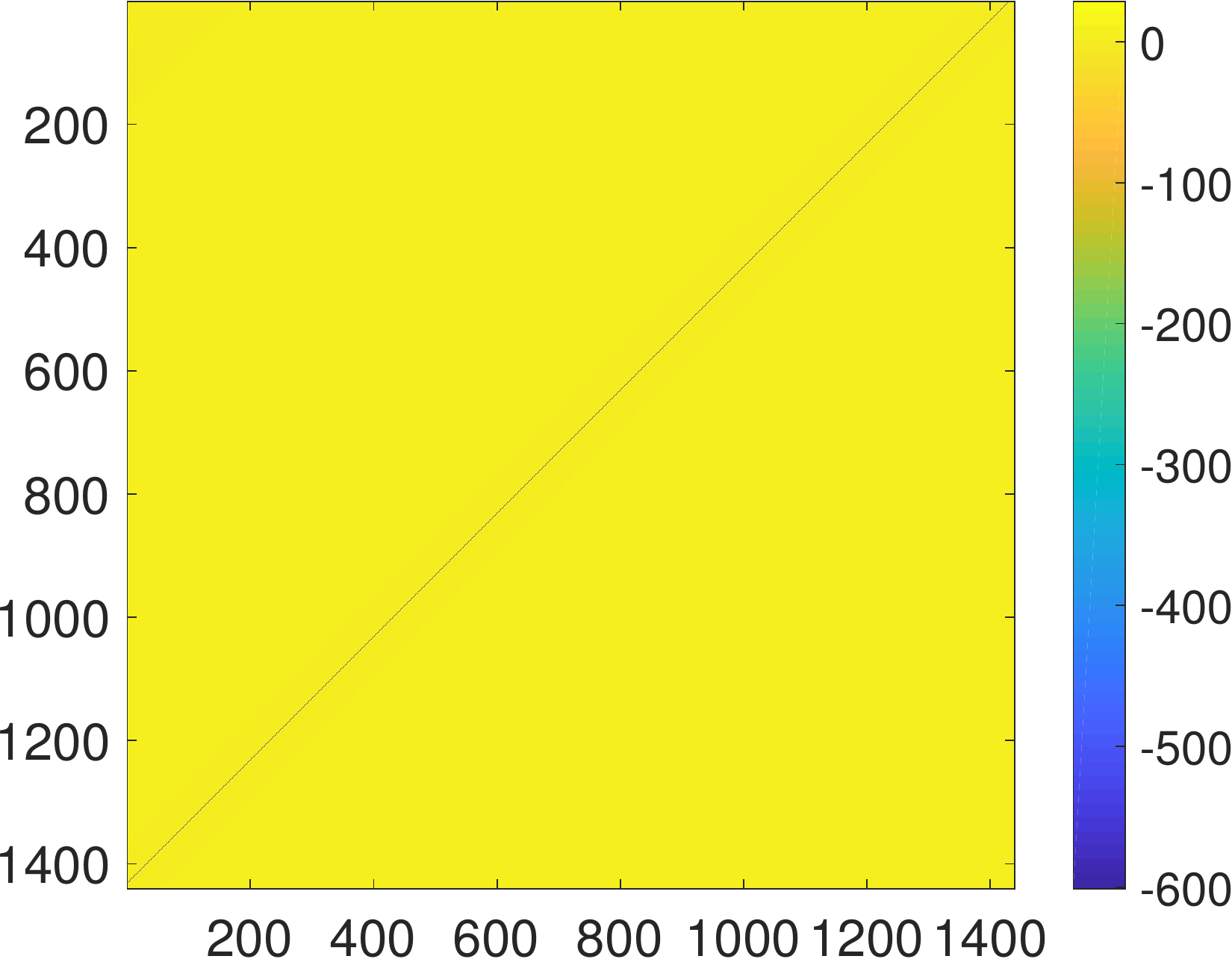} & \includegraphics[scale=0.3]{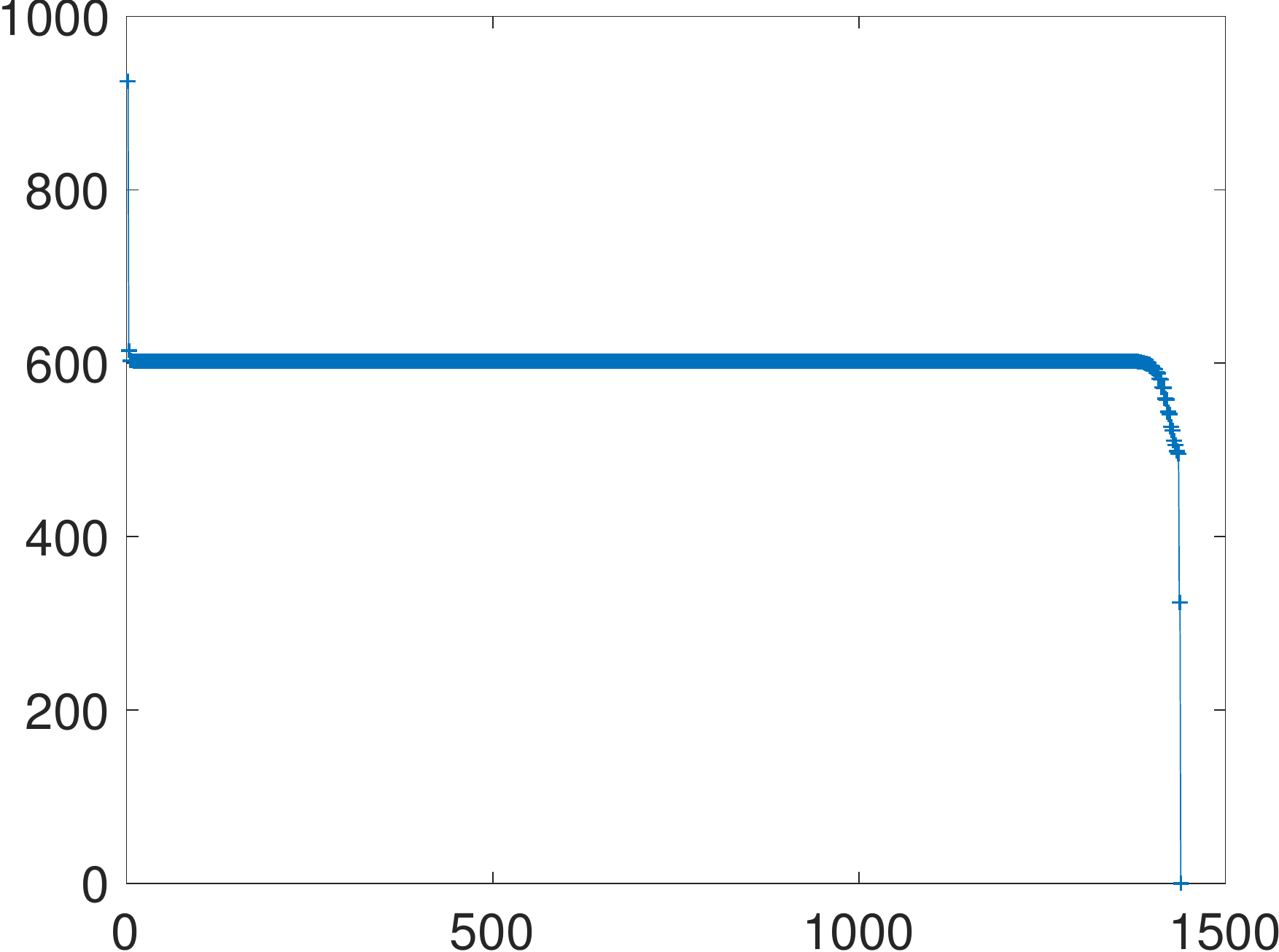}
  \end{tabular}
  \caption{Comparison between $b(t)=0$ and $b(t) \sim \sin(d t) + \ldots$.  The top row is for
    $b(z)=0$ and the bottom row is $b(t) \sim \sin(d t) + \ldots$. Within each row, the left plot is
    the matrix in \eqref{eq:ls} and the right is its singular values. Notice that the leading
    term $\sin(dt)$ in $b(t)$ introduces a dominant anti-diagonal in the matrix, ensuring that it
    has numerical rank $2d$.  }
  \label{fig:bMS}
\end{figure}

Figure \ref{fig:bMS} illustrates the difference between $b(t)=0$ and $b(t)\sim\sin(dt)+\ldots$
for the Fermi-Dirac operator (the last example in Section \ref{sec:res}) at $\beta=100$. Notice that
the leading term $\sin(dt)$ in $b(t)$ introduces a dominant anti-diagonal in the matrix of
\eqref{eq:ls}, ensuring that it has numerical rank $2d$.

\subsection{Implementation}\label{sec:imp}

To implement this algorithm numerically, we need to take care several issues.

\begin{itemize}
\item The computation \eqref{eq:Fourier} requires the Fourier transform of
  $g(t)=(1-a^2(t)-b^2(t))^{-1}$ for $t\in[-\pi,\pi]$.  When $1-a^2(t)-b^2(t)$ is close to zero,
  $g(t)$ is near singular and hence it is hard for numerical quadrature. In practice, we make sure
  that $\|a\|_\infty$ and $\|b\|_\infty$ are bounded by 1/3.

  
\item 
  To compute the Fourier coefficients $\{\hg_k\}$, choose an even $N_s$ and define for
  $n=0,\ldots,N_s-1$ the point $t_n = \exp\left(i \frac{2\pi n}{N_s}\right)$ on the unit circle.
  Using samples $\{g(t_n)\}$ at the points $\{t_n\}$ corresponds to approximating
  \eqref{eq:integral} with the trapezoidal rule. The trapezoidal rule is exponentially convergent
  for smooth functions when $N_s$ is sufficient large. In the current setting since
  $1-a^2(t)-b^2(t)$ is bounded well away from zero, $g(t)$ does not exhibit singular behaviors. As a
  result, the highest non-trivial frequency in $g(t)$ is on the same order of the highest
  non-trivial frequency in $a(t)$ and $b(t)$. Therefore, by setting $N_s$ to be about $40$ times $d$
  in practice, we ensure that the trapezoidal rule is exponentially accurate.  Applying the fast
  Fourier transform to $\{g(t_n)\}$ gives the Fourier coefficients $\{\hg_k\}$.
  
\item The semi-infinite matrix in \eqref{eq:ls}. In the implementation, we only pick the first
  $l$ rows of this semi-infinite matrix and define
  \begin{equation}
  H \equiv
  \begin{bmatrix}
    \hg_{-1} & \hg_{-2} & \cdots & \hg_{-(2d+1)} \\
    \hg_{-2} & \hg_{-3} & \cdots & \hg_{-(2d+2)} \\
    \vdots & \vdots & \vdots & \vdots\\
    \hg_{-l} & \hg_{-(l+1)} & \cdots & \hg_{-(2d+l)}
  \end{bmatrix}
  \label{eq:H}
  \end{equation}
  with $l\ge 2d+1$. In practice setting $l=2d+2$ seems to be sufficient.

\item
  The computation of the vector $m$. The most straightforward way is to compute the singular value
  decomposition (SVD) of $H$ in \eqref{eq:H} and take $m$ to be the last column of the $V$ matrix,
  which unfortunately has $O(d^3)$ time complexity.
  
  This complexity can be improved based on the following observation. Let $s_1,\ldots,s_{2d+1}$ be
  the singular values of $H$. Numerically, our choice of $b(t)$ leads to a large gap between
  $s_{2d}$ and $s_{2d+1}$ that is actually proportional to $s_1$. As a result, we propose the
  following iterative procedure
  \begin{equation}
    m \Leftarrow \text{normalize} \left( \left( \eps I + H^T H \right)^{-1} m \right),
    \label{eq:miter}
  \end{equation}
  where $\eps$ is small positive constant. The linear system solve within each iteration is done
  with the conjugate gradient (CG) method and the iteration stops when the difference between the
  new and old $m$ is less than the machine accuracy. Due to the large spectral gap of the matrix
  $H$, the inner CG method typically converges within a constant number of iterations and the outer
  iteration stops in 3-4 iterations. Since the matrix $\eps I + H^T H$ is of size
  $(2d+1)\times(2d+1)$, the empirical cost for computing $m$ is $O(d^2)$. Combined with the
  $O(d^2\log d)$ cost of extracting the phase factors \eqref{eq:pr}, the overall cost is $O(d^2\log
  d)$.
  
  The complexity for computing $m$ can be further reduced by observing that $H$ is a Hankel matrix.
  Therefore, the matrix-vector multiplications of $H$ and $H^T$ can be accelerated to the $O(d\log
  d)$ cost via the fast Fourier transform (FFT). This improvement does not impact the overall
  computational cost, since the construction of the phase factors from $p(z)$ (Section
  \ref{sec:frompz}) already takes $O(d^2 \log d)$ steps. On the other hand, since there is no need
  to store the full $H$ matrix, the memory cost is reduced to $O(d)$, allowing for working with very
  large $d$ values.
\end{itemize}  
  
A few remarks are in order here.
\begin{itemize}
\item The symmetric phase factors obtained in \cite{dong2021efficient} correspond to $b(t)$=0, which
  helps the optimization approach. In the current algorithm, the choice of $b(t) = b_d \sin(dt) + \ldots$
  helps the direct factorization method and the resulting phase factors are non-symmetric.
\item Whether symmetric or non-symmetric phase factors are preferred in practice is not clear. The
  main part of implementing QSVT on quantum circuits as in \eqref{eq:Ut} is actually related to the
  terms $e^{i t X}$, i.e., the implementation of the circuit $U_A$. The actual choice of the phase
  factors might very well depend on the circuit architecture.
\end{itemize}

\section{Results}\label{sec:res}



\subsection{Setup}
The algorithm is implemented with the standard double precision arithmetics. All numerical results
are obtained on a laptop with a 2.6 GHz 6-Core Intel Core i7 CPU. The computation of the vector $m$
is performed via the iteration \eqref{eq:miter}. The complexity is quadratic in terms of the degree
$d$ and the actual computation typically finishes within a couple of minutes.


As mentioned in Section \ref{sec:intro}, when the target function $f(x)$ is not a polynomial, the
first step is to construct an accurate polynomial approximation $a(x)$. Since polynomial
approximation in $x$ is equivalent to trigonometric approximation in $t$, this task is performed the
$t$ space. Let us introduce an equally spaced grid $t_n = \exp\left(i \frac{2\pi n}{N_s}\right)$ on
the unit circle, where the grid size $N_s$ is taken in practice to be $40$ times $d$ as before.
\begin{itemize}
\item First, the values of $f(t)$ at $\{t_n\}$ are computed.
\item 
  After applying fast Fourier transforms to $\{f(t_n)\}$, we identify a frequency $d$ such that all
  Fourier coefficients above frequency $d$ are below a threshold multiplied by the maximum Fourier
  coefficient in absolute value. In the experiments, the threshold is chosen to be around $10^{-12}$
  since this is right above the accuracy of the QSP algorithm in double precision arithmetics
  \cite{dong2021efficient} and further improvement below this threshold is not necessary. Here $d$
  is enforced to have the same parity as $f(x)$.
\item The Fourier coefficients above frequency $d$ are
  then set to zero and Fourier transform back gives the desired trigonometric approximation to
  $f(t)$ in the $t$ space. The final function $a(t)$ is also scaled to have infinity norm equal to
  $0.3$.
\end{itemize}

Regarding the choice of $b(t)$, simply setting $b(t)=0.4\cdot\sin(dt)$ suffices for in the examples
presented below. However, in principle, one might need to choose to $b(t)$ according to \eqref{eq:b}
in order to avoid identical roots.

The polynomial coefficient vector $m$ is computed with the iterative procedure \eqref{eq:miter}. The
error of the constructed phase factors is measured in the relative $L_\infty$ norm. More precisely,
we compute a function $\tilde{p}(x)$ following \eqref{eq:U} using the constructed phase factors
$\Phi=(\phi_0,\ldots,\phi_d)$. With $\tilde{p}(x)$ as an approximation to $p(x)$, the error of the
phase factor computation is estimated with
\begin{equation}
  \frac{ \|\Re(\tilde{p}(x))-a(x)\|_\infty }{ \|a(x)\|_\infty },
  \label{eq:errest}
\end{equation}
over the interval $[-1,1]$.


\subsection{Examples}

In what follows, we present the numerical results for four examples: Hamiltonian simulation,
eigenstate filtering, matrix inversion, and Fermi-Dirac operator. Among them, the first three
examples are also studied numerically in the paper \cite{dong2021efficient}, arguably the most
complete numerical study on the phase factor computation. In each example, we have chosen the
parameters to be on par or harder compared with those used in \cite{dong2021efficient}, so that the
actual instances have at least the same level of difficulty. Overall, our algorithm exhibits similar
accuracy but short wall clock time when compared with \cite{dong2021efficient}. The longest sequence
reported below consists of more than $50000$ phase factors.

\subsubsection{Hamiltonian simulation} 
\begin{figure}[h!]
  \centering
  \begin{tabular}{cc}
    \includegraphics[scale=0.30]{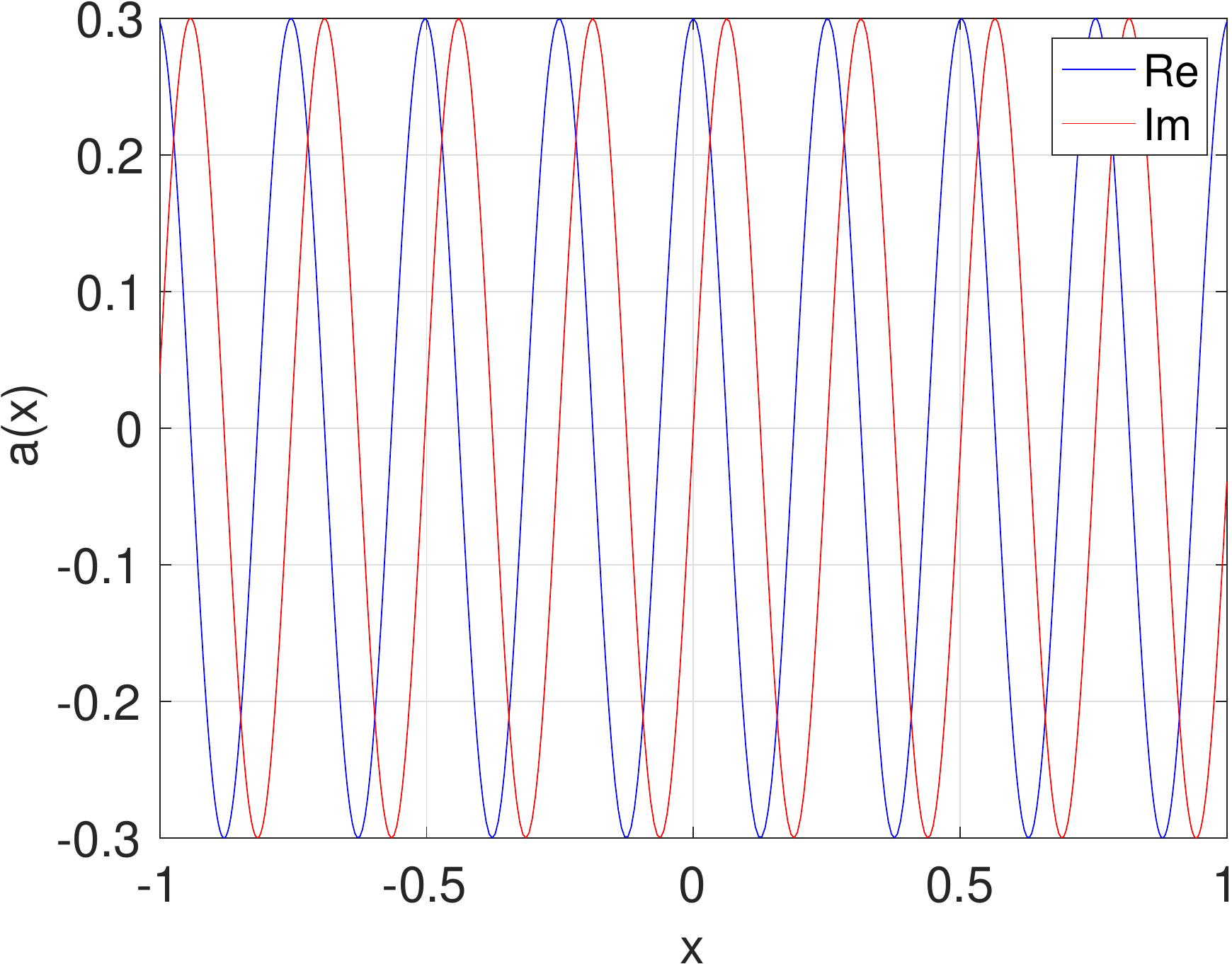} & \includegraphics[scale=0.30]{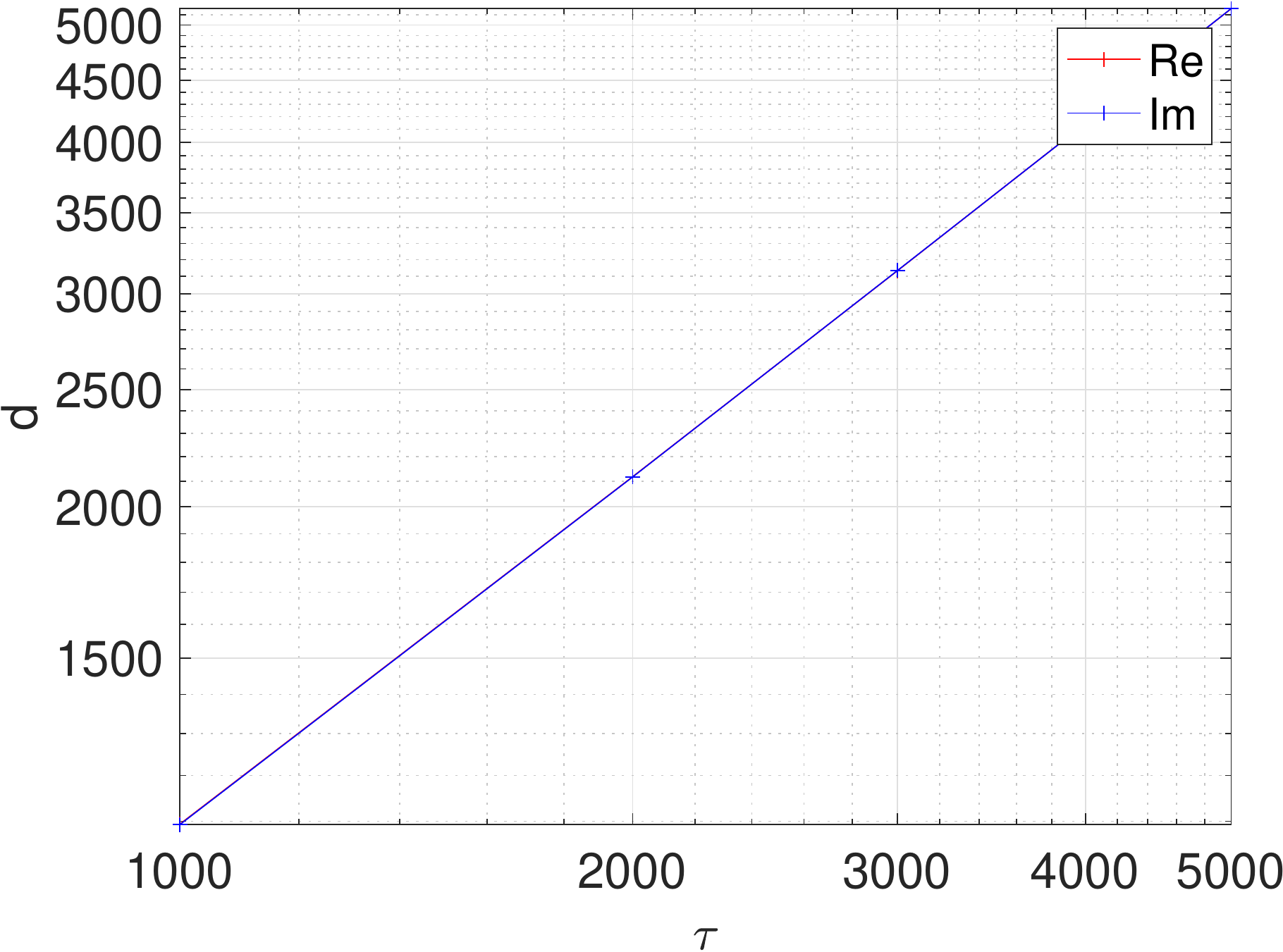} \\
    (a) & (b)\\
    \includegraphics[scale=0.30]{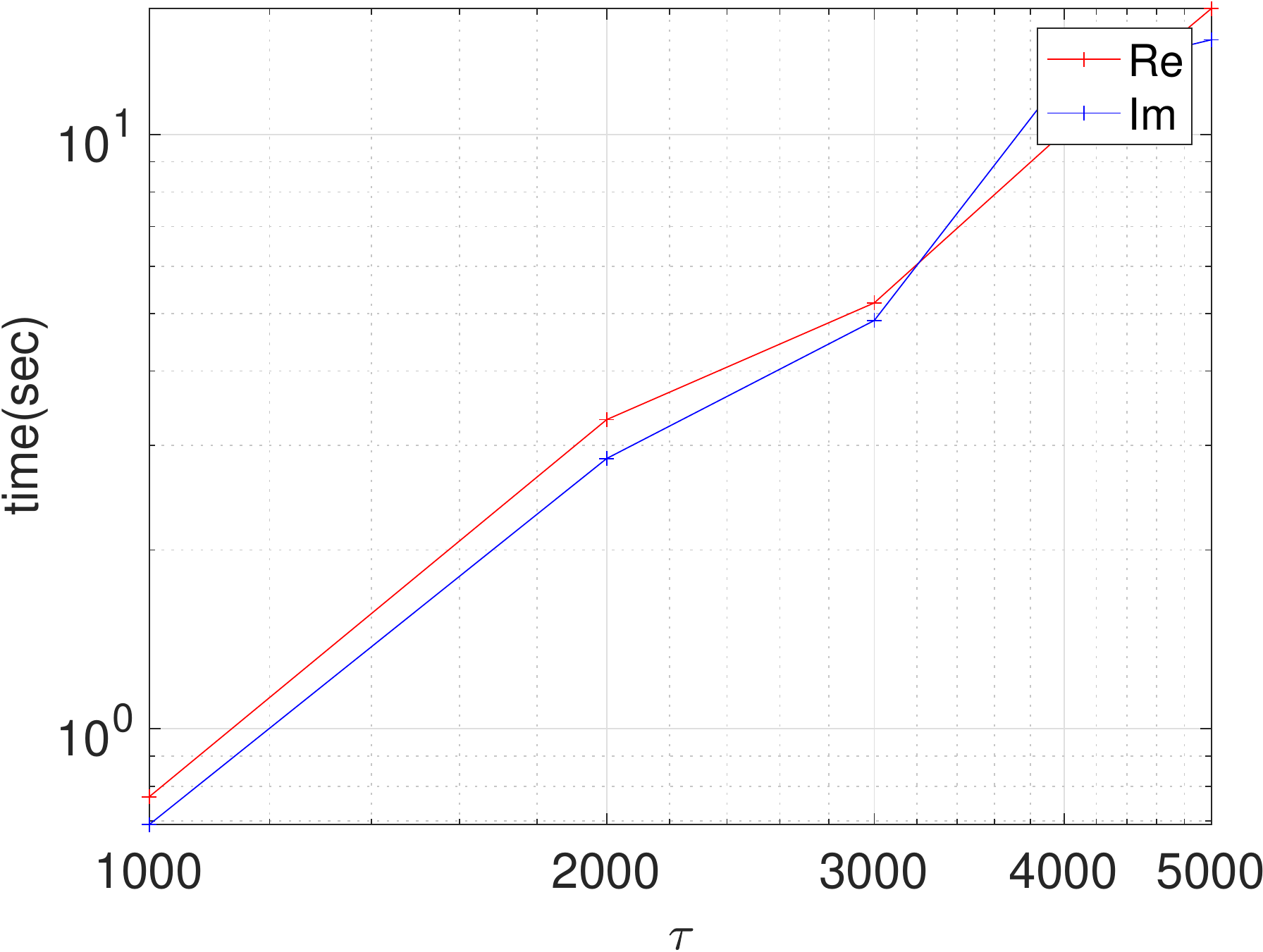} & \includegraphics[scale=0.30]{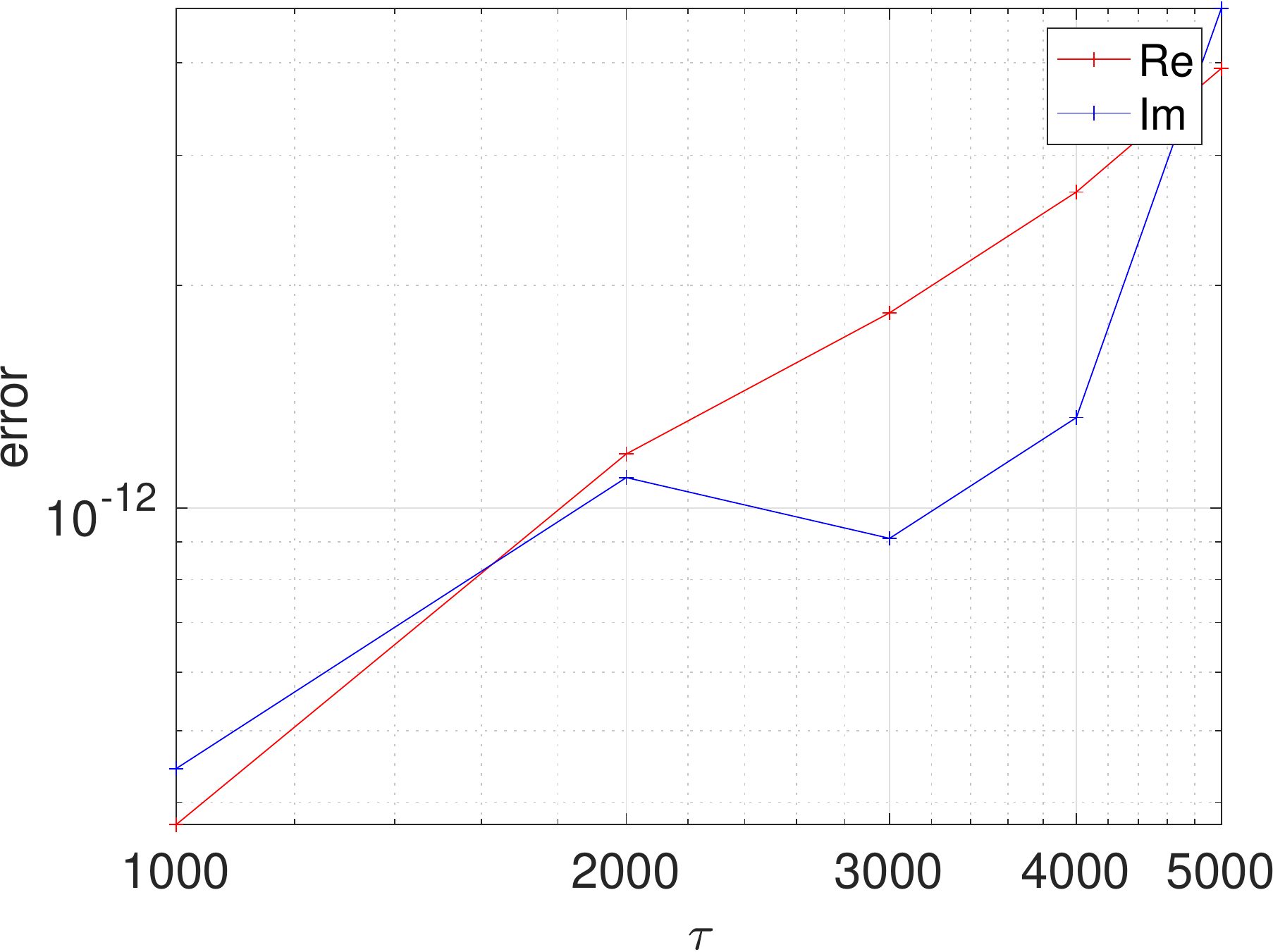}\\
    (c) & (d)
  \end{tabular}
  \caption{Hamiltonian simulation.  (a) $a_\Re(x)$ and $a_\Im(x)$ for a smaller $\tau=25$. (b)
    degree $d$ of the polynomial approximation as a function of $\tau$. (c) The total phase factor
    construction time in seconds as a function of $\tau$. (d) The relative $L_\infty$ norm error
    \eqref{eq:errest} as a function of $\tau$.}
  \label{fig:ham}
\end{figure}

Assume that the Hamiltonian $H$ satisfies $\|H\|_2 \le 1$.
Hamiltonian simulation for a period of time $\tau$ boils down to the quantum signal processing
problem for $f(x) = e^{-i\tau x}$. The even and odd parts are
\[
\Re(f(x)) = \cos(\tau x), \quad \Im(f(x)) = \sin(\tau x).
\]
In terms of the variable $t$, 
\[
\Re(f(t)) = \cos(\tau \cos(t)), \quad \Im(f(t)) = \sin(\tau \cos(t)).
\]
Since both functions are not polynomials, we first use the procedure described above to compute
trigonometric polynomial approximations $a_\Re(t) \approx 0.3 \cdot \Re(f(t))$ and $a_\Im(t) \approx
0.3 \cdot \Im(f(t))$, both scaled so that $L_\infty$ norm is below $1/3$. $a_\Re(x)$ is even in $x$
with even degree $d_\Re$ while $a_\Im(x)$ is odd in $x$ with odd degree $d_\Im$. We perform the
tests with $\tau=1000,2000,3000,4000,5000$ and the results are summarized in Figure \ref{fig:ham}.

\subsubsection{Eigenstate filtering} 
\begin{figure}[h!]
  \centering
  \begin{tabular}{cc}
    \includegraphics[scale=0.30]{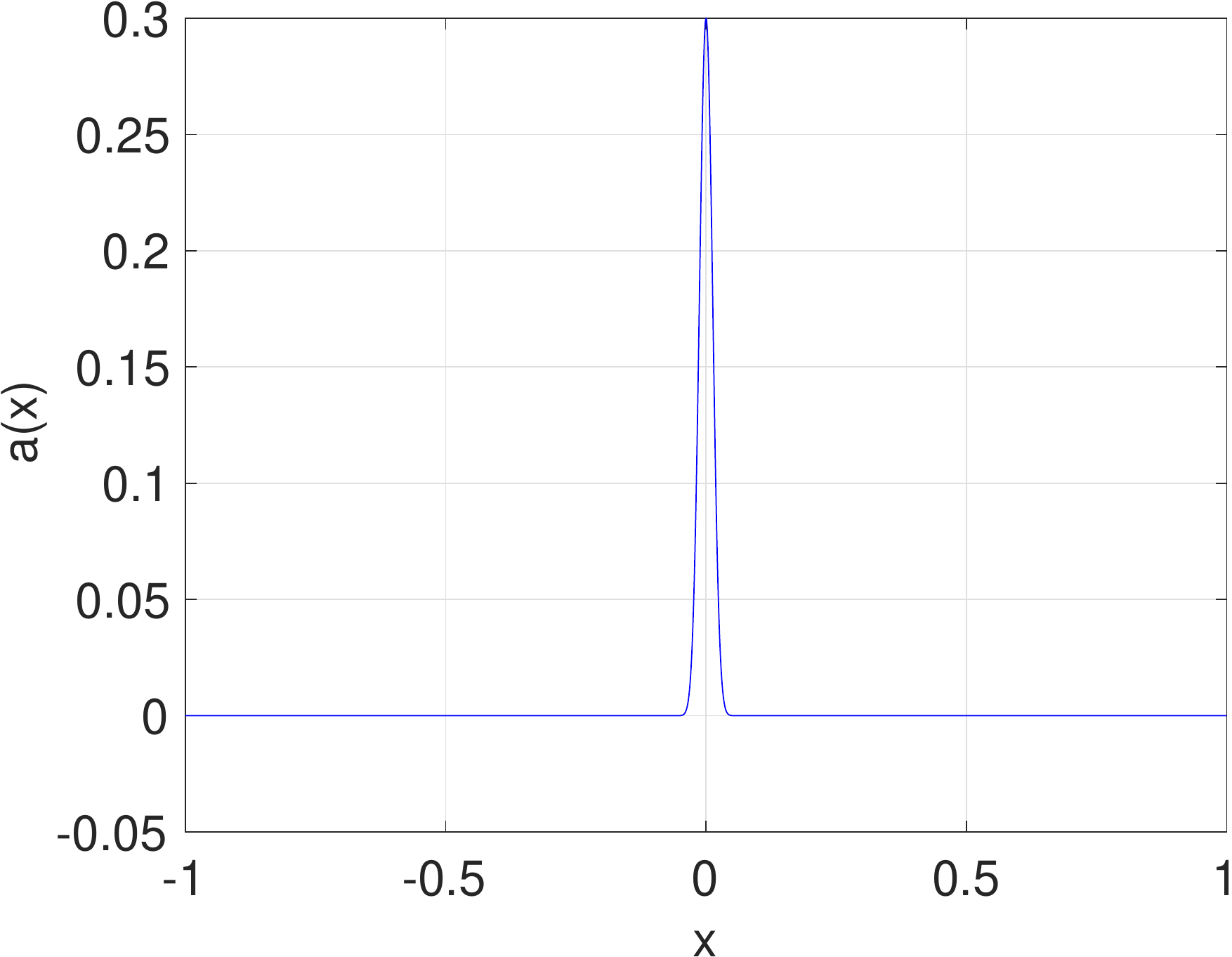} & \includegraphics[scale=0.30]{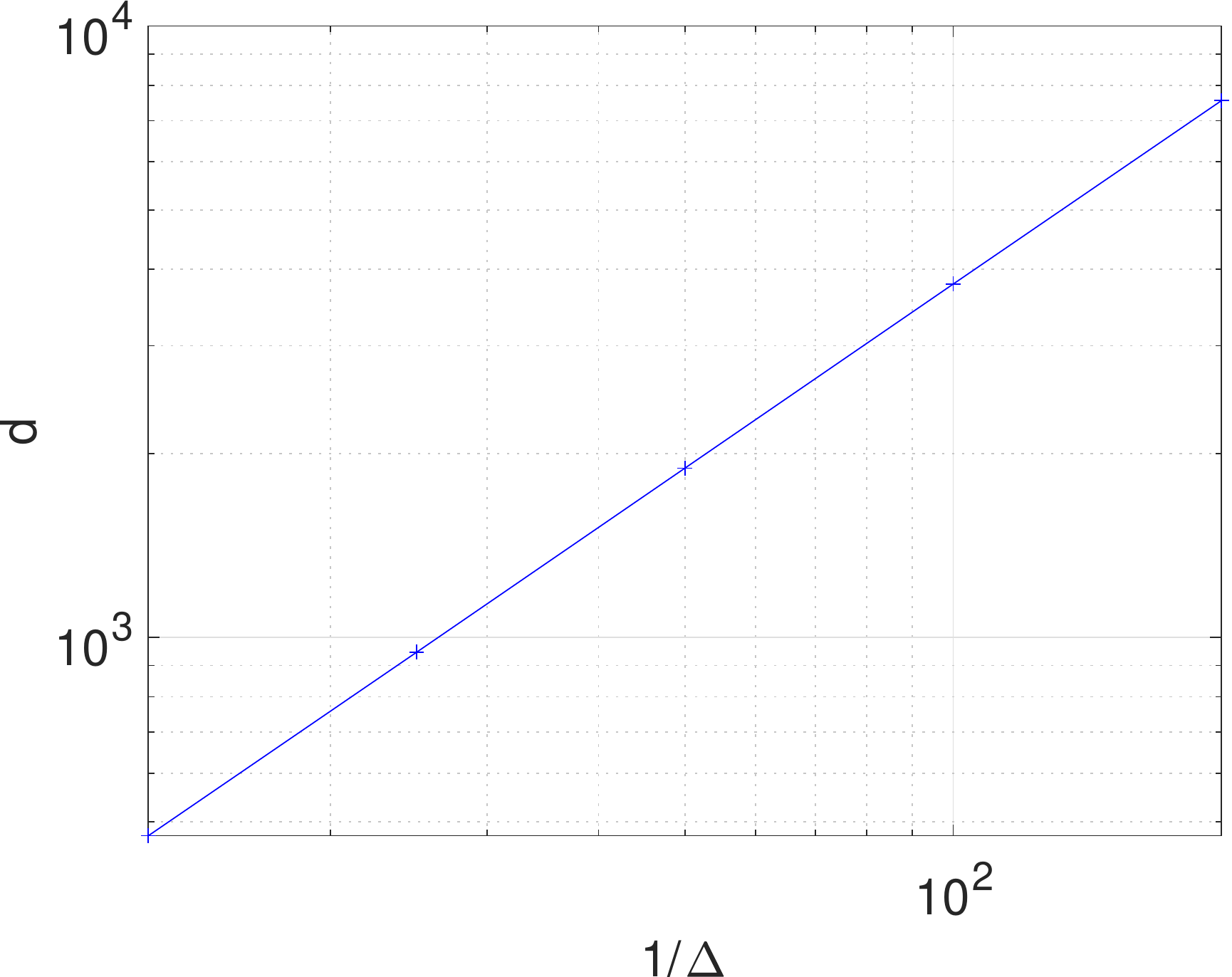}\\
    (a) & (b)\\
    \includegraphics[scale=0.30]{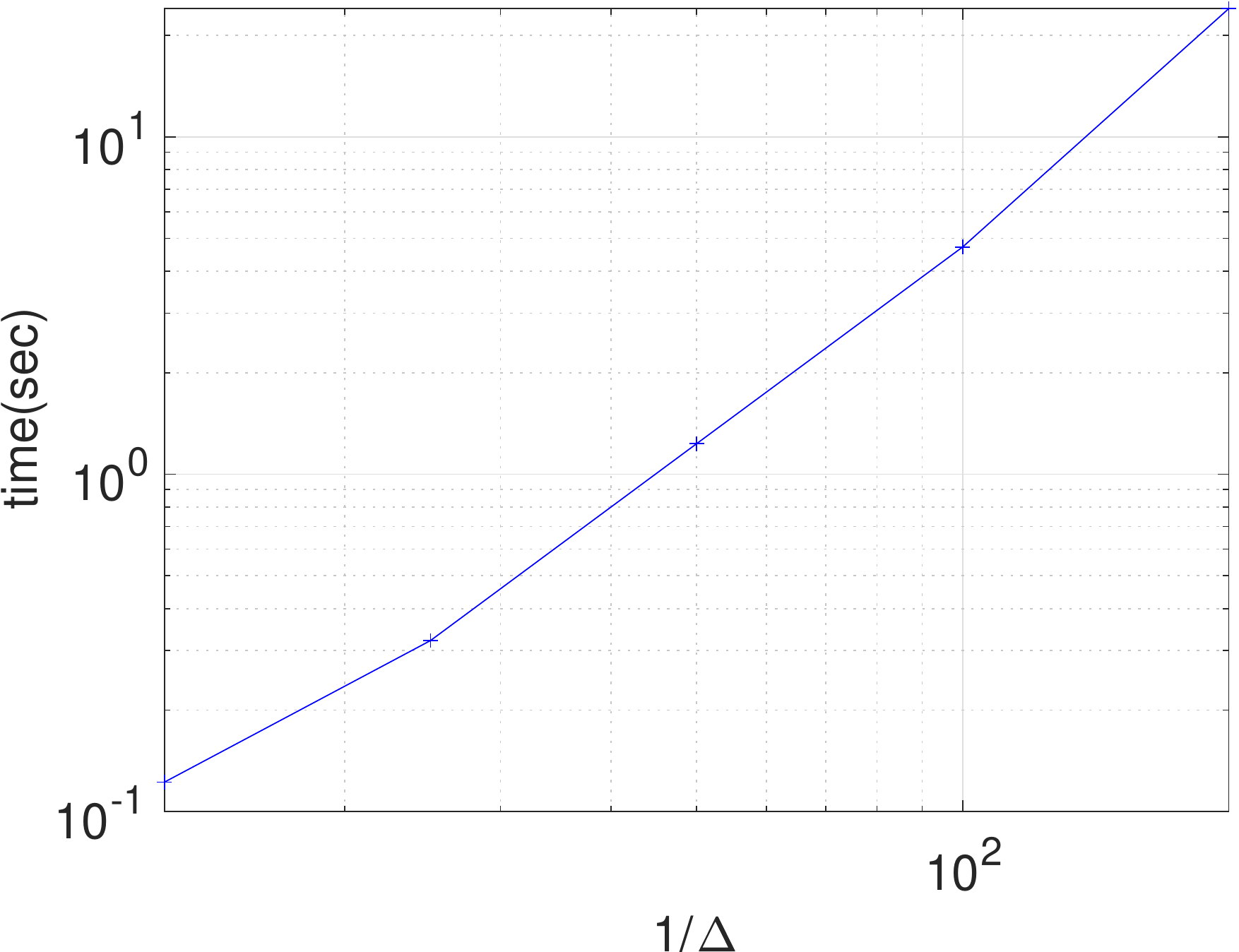} & \includegraphics[scale=0.30]{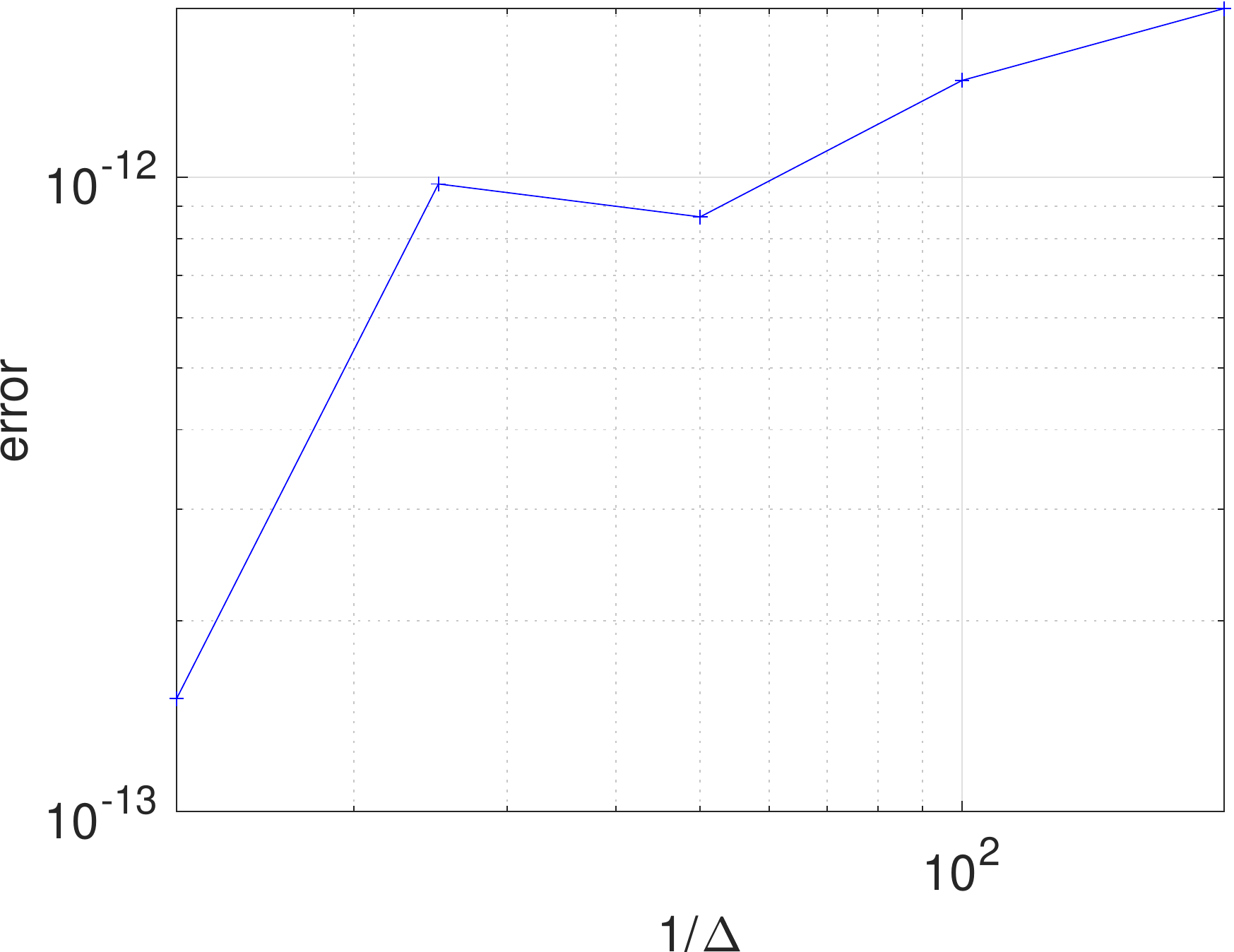}\\
    (c) & (d)
  \end{tabular}
  \caption{Eigenstate filtering. (a) $a(x)$ for $\Delta=0.08$. (b) The degree $d$ of the polynomial
    approximation as a function of $1/\Delta$.  (c) The total phase factor construction time in
    seconds as a function of $1/\Delta$. (d) The relative $L_\infty$ norm error \eqref{eq:errest} as a
    function of $1/\Delta$.}
  \label{fig:filter}
\end{figure}

For a fixed gap $\Delta$, we follow \cite{dong2021efficient} and
consider the filtering function centered at the origin
\[
f(x) = \frac{T_k\left( -1 + 2 \frac{x^2-\Delta^2}{1-\Delta^2} \right)}{T_k\left( -1 + 2
  \frac{-\Delta^2}{1-\Delta^2} \right)},
\]
where $T_k$ is the Chebyshev polynomial. The parameter $k$ is set to be $20/\Delta$ so that the
$f(x)$ is negligible outside the $\Delta$ neighborhood of the origin. In terms of variable $t$, the
function
\[
f(t) = \frac{T_k\left( -1 + 2 \frac{\cos(t)^2-\Delta^2}{1-\Delta^2} \right)}{T_k\left( -1 + 2 \frac{-\Delta^2}{1-\Delta^2} \right)}.
\]
Since $f(t)$ is already a trigonometric polynomial, we simply set $a(t)=0.3 \cdot f(t)$. The tests
are performed with $\Delta=0.08, 0.04, 0.02, 0.01, 0.005$ and the results are summarized in Figure
\ref{fig:filter}.

\subsubsection{Matrix inversion}
\begin{figure}[h!]
  \centering
  \begin{tabular}{cc}
    \includegraphics[scale=0.30]{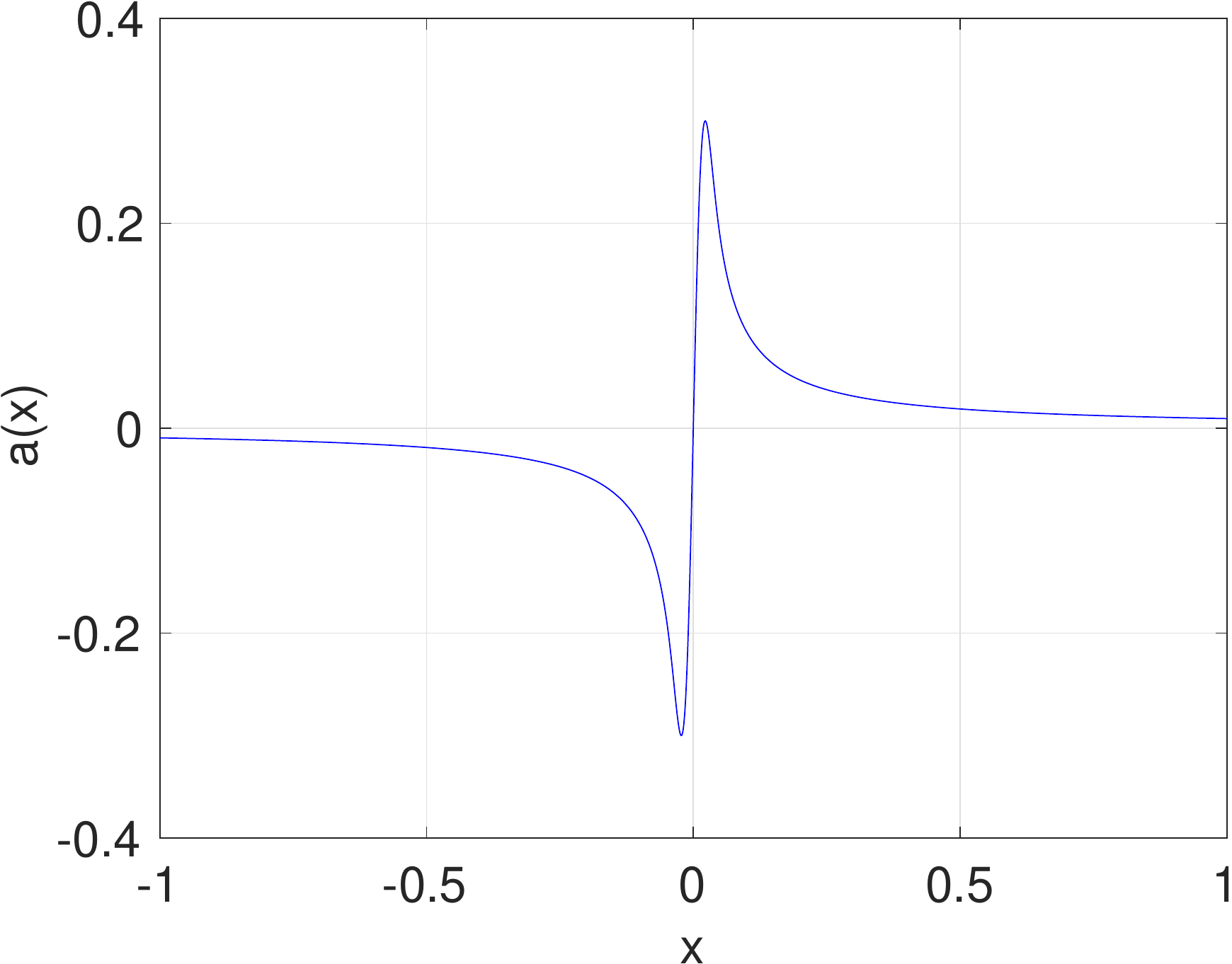} & \includegraphics[scale=0.30]{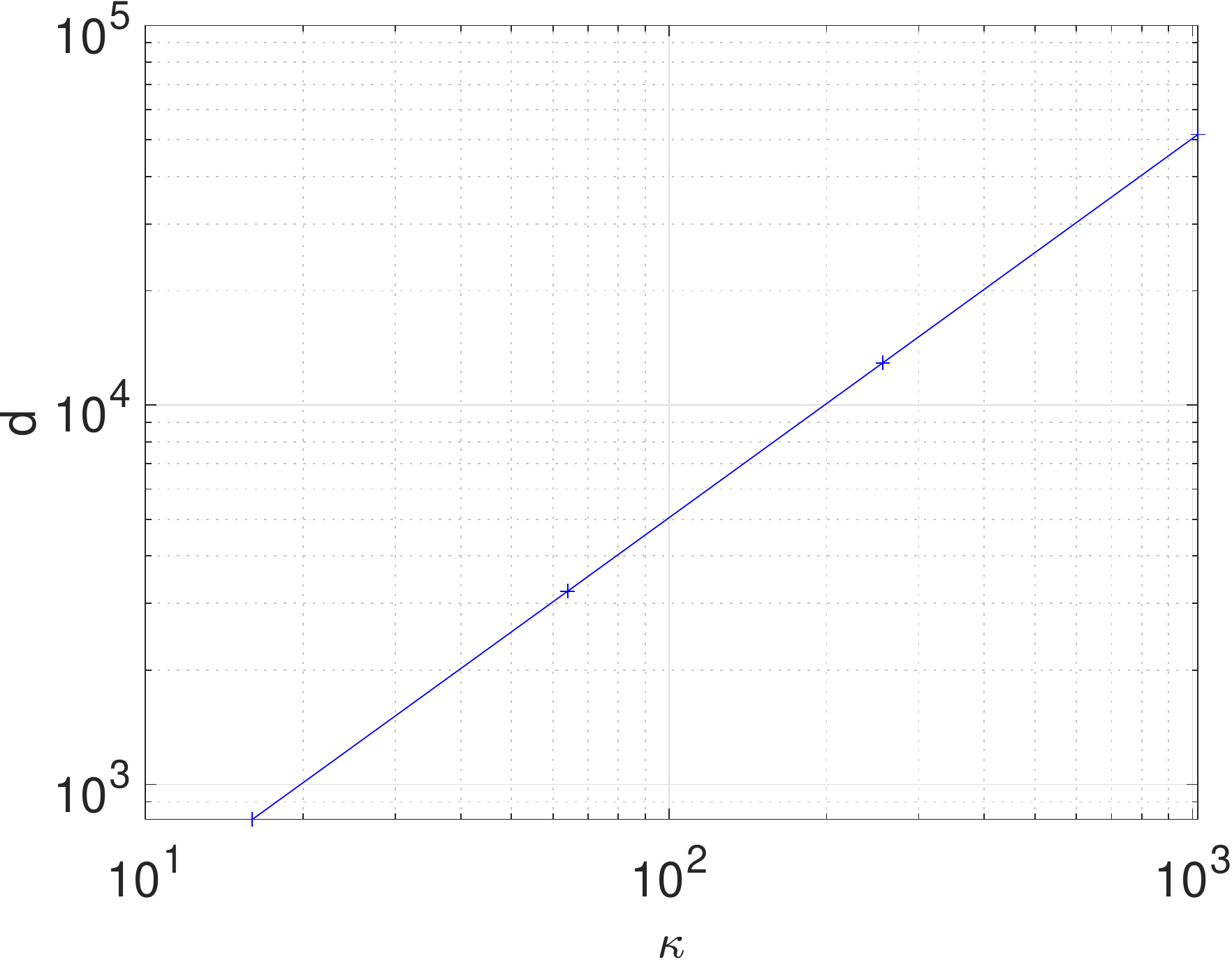} \\
    (a) & (b) \\
    \includegraphics[scale=0.30]{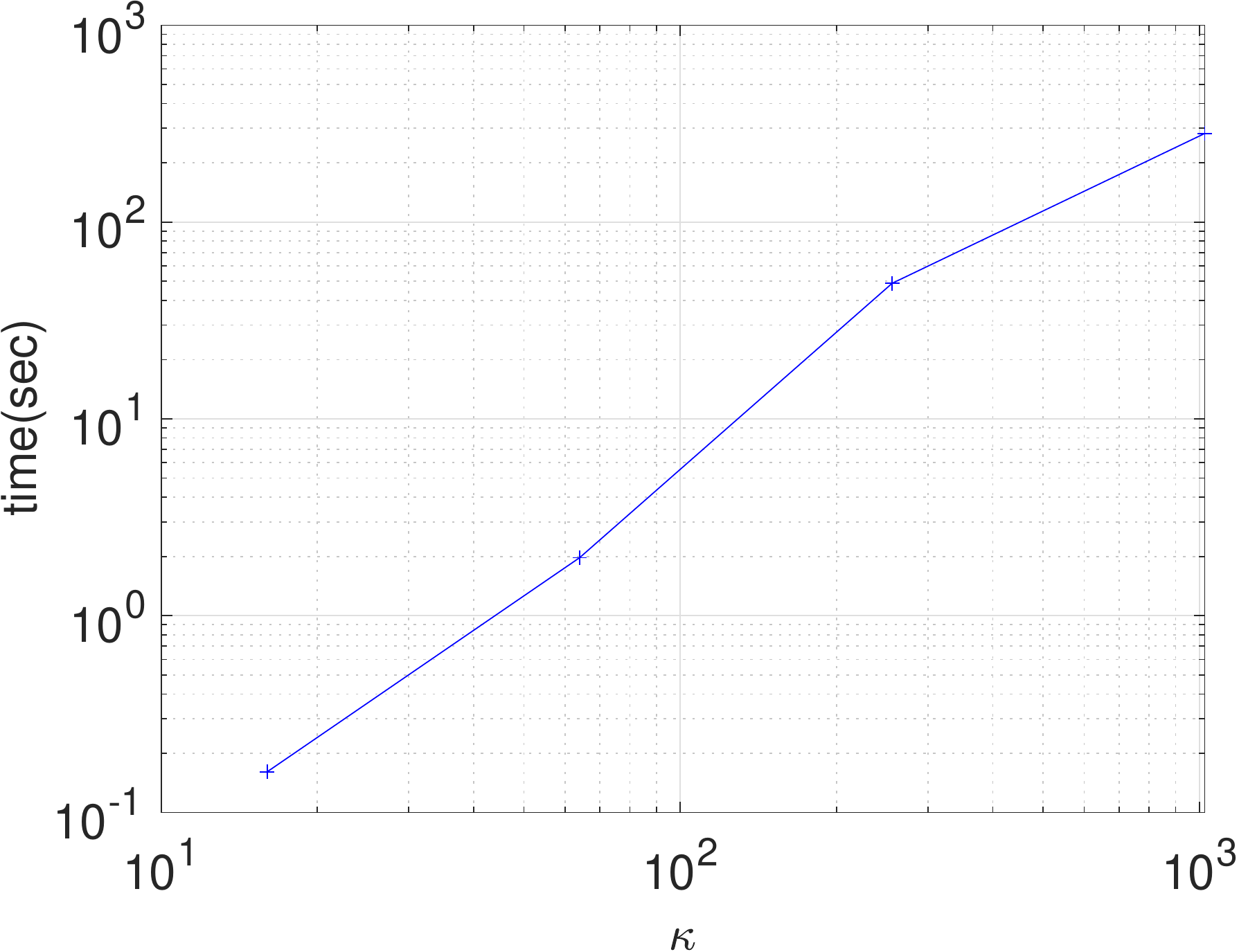} & \includegraphics[scale=0.30]{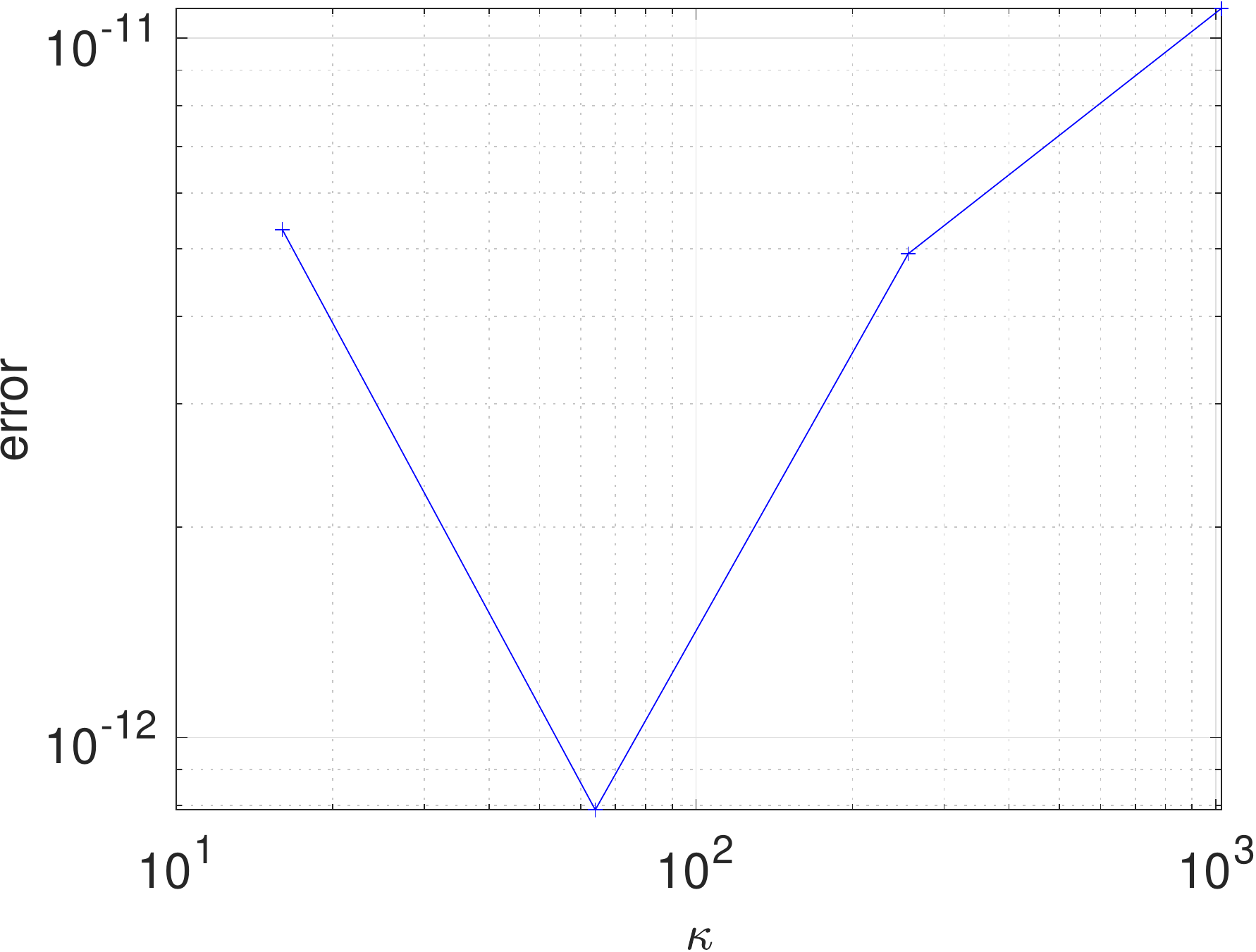}\\
    (c) & (d)
  \end{tabular}
  \caption{Matrix inversion. (a) $a(x)$ for $\kappa=10$. (b) The degree $d$ of the polynomial
    approximation as a function of $\kappa$.  (c) The total phase factor construction time in
    seconds as a function of $\kappa$. (d) The relative $L_\infty$ norm error \eqref{eq:errest} as
    a function of $\kappa$.}
  \label{fig:inv}
\end{figure}

We consider the inversion of the matrices with spectrum resided in
$D_\kappa=[-1,-1/\kappa]\cup[1/\kappa,1]$, where $\kappa$ is the condition number. The QSP problem
here amounts to approximating the function $1/x$ over $D_\kappa$. We choose
\[
f(x) = \frac{1 - e^{- (5\kappa x)^2}}{x},
\]
where the difference between $f(x)$ and $1/x$ over $D_\kappa$ is negligible under the double
precision arithmetics. In the $t$ variable, this is
\[
f(t) = \frac{1 - e^{- (5\kappa \cos(t))^2}}{\cos(t)}.
\]
The procedure mentioned above is used to compute a trigonometric approximation $a(t)$ to $f(t)$ (up
to a constant factor) with $\|a\|_\infty=0.3$.  The tests are performed with $\kappa=16,64,256,1024$
and the results are summarized in Figure \ref{fig:inv}. The longest sequence for $\kappa=1024$ has
more than $50000$ phase factors.

\subsubsection{Fermi-Dirac operator}
\begin{figure}[h!]
  \centering
  \begin{tabular}{cc}
    \includegraphics[scale=0.30]{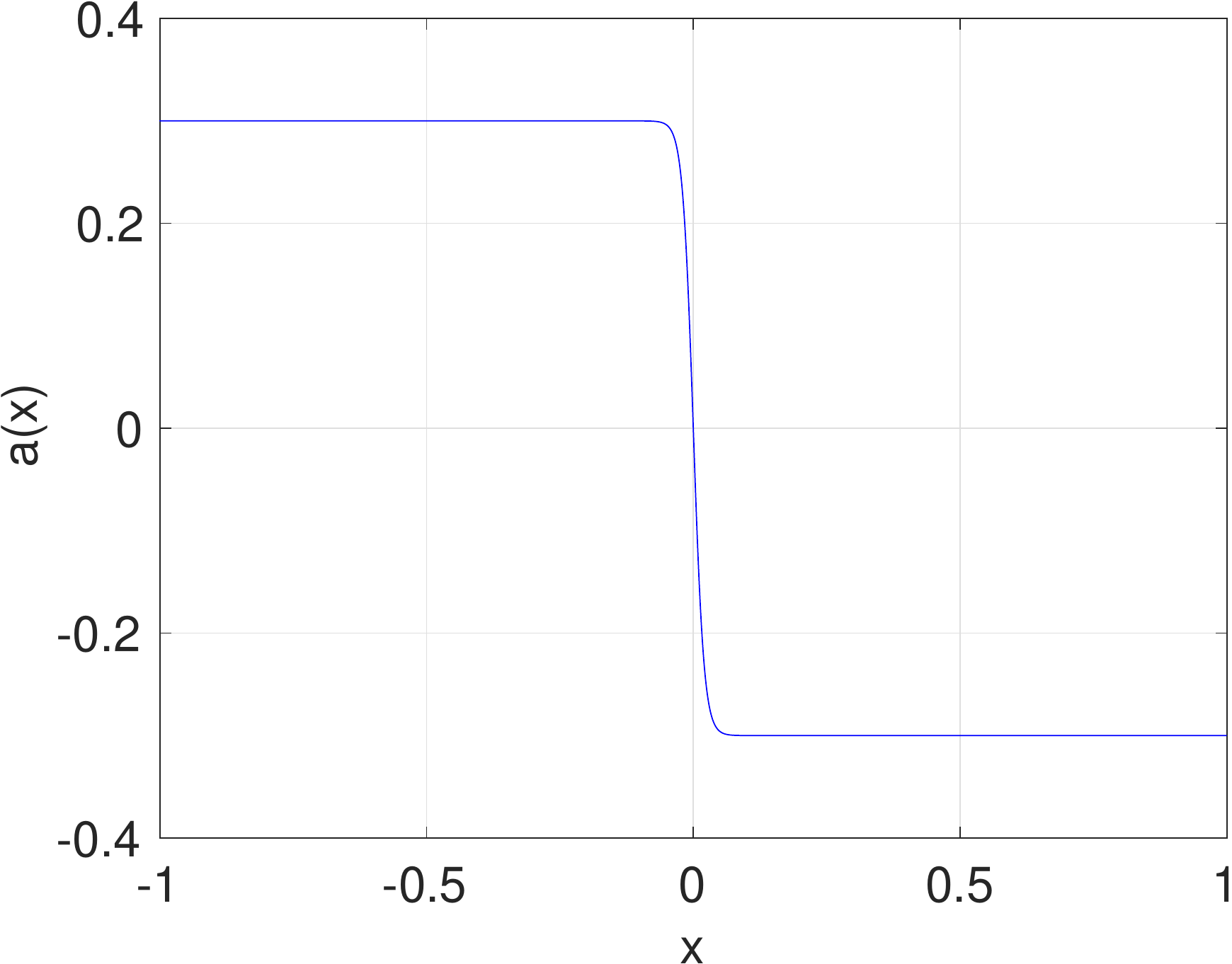} & \includegraphics[scale=0.30]{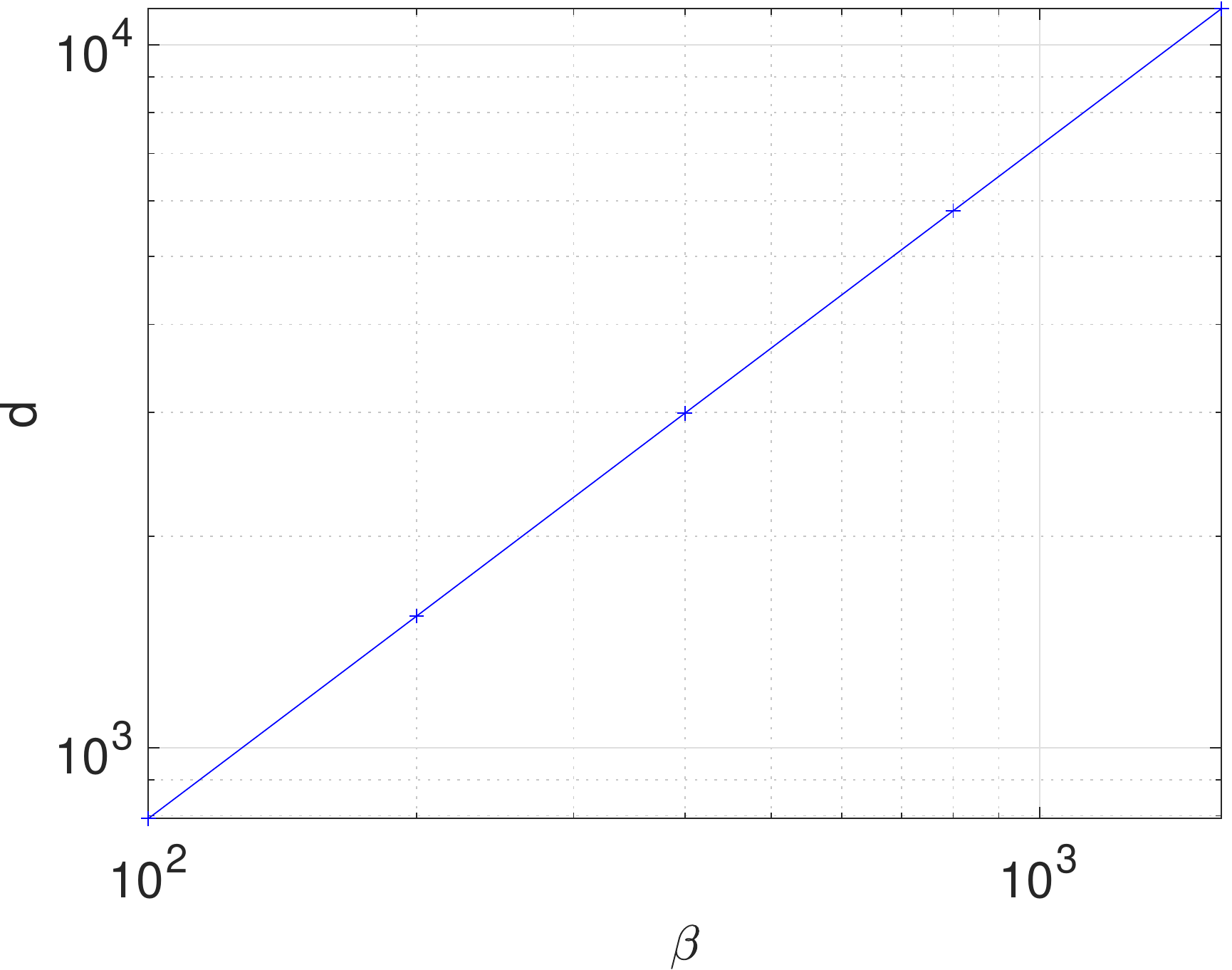}\\
    (a) & (b)\\
    \includegraphics[scale=0.30]{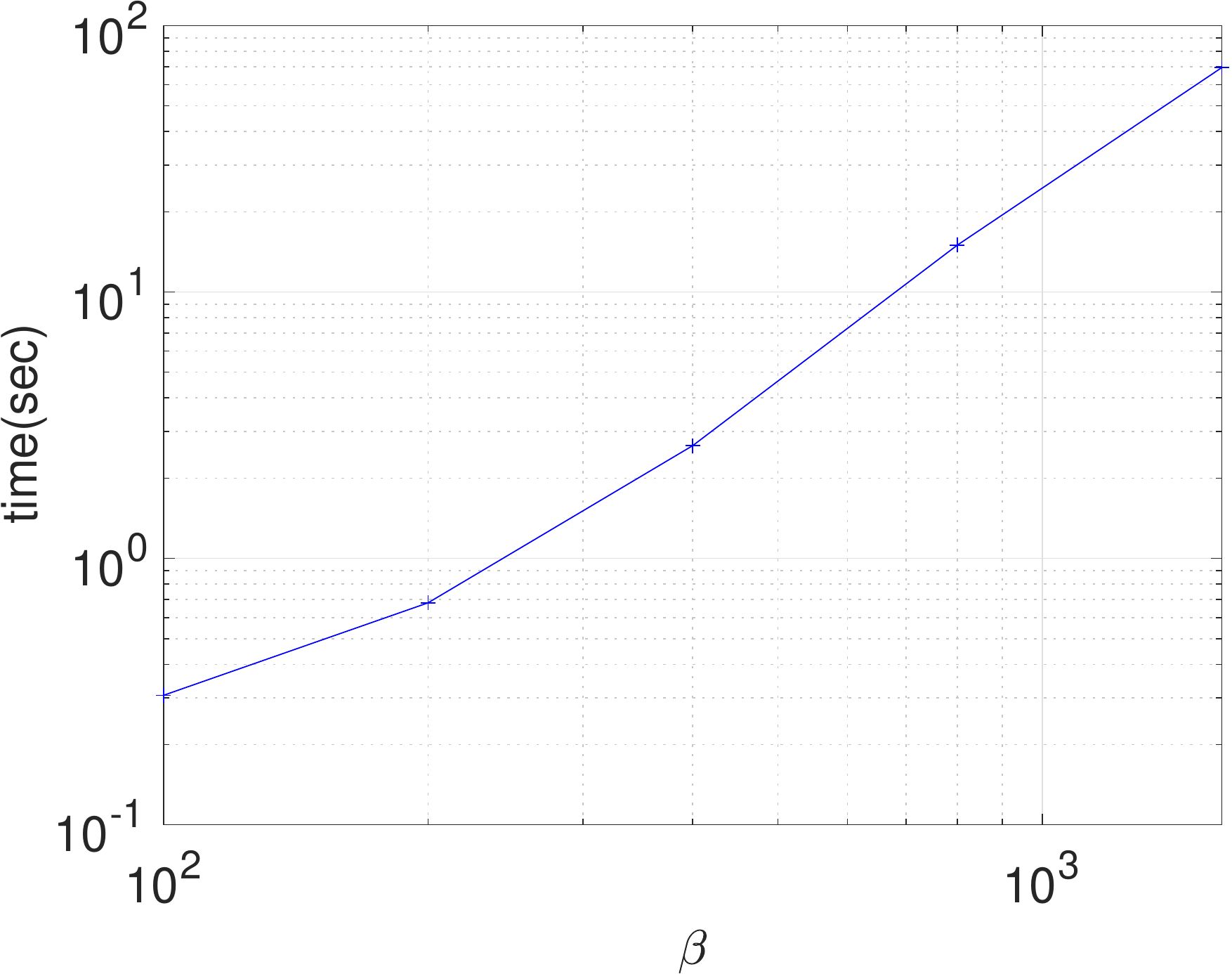} & \includegraphics[scale=0.30]{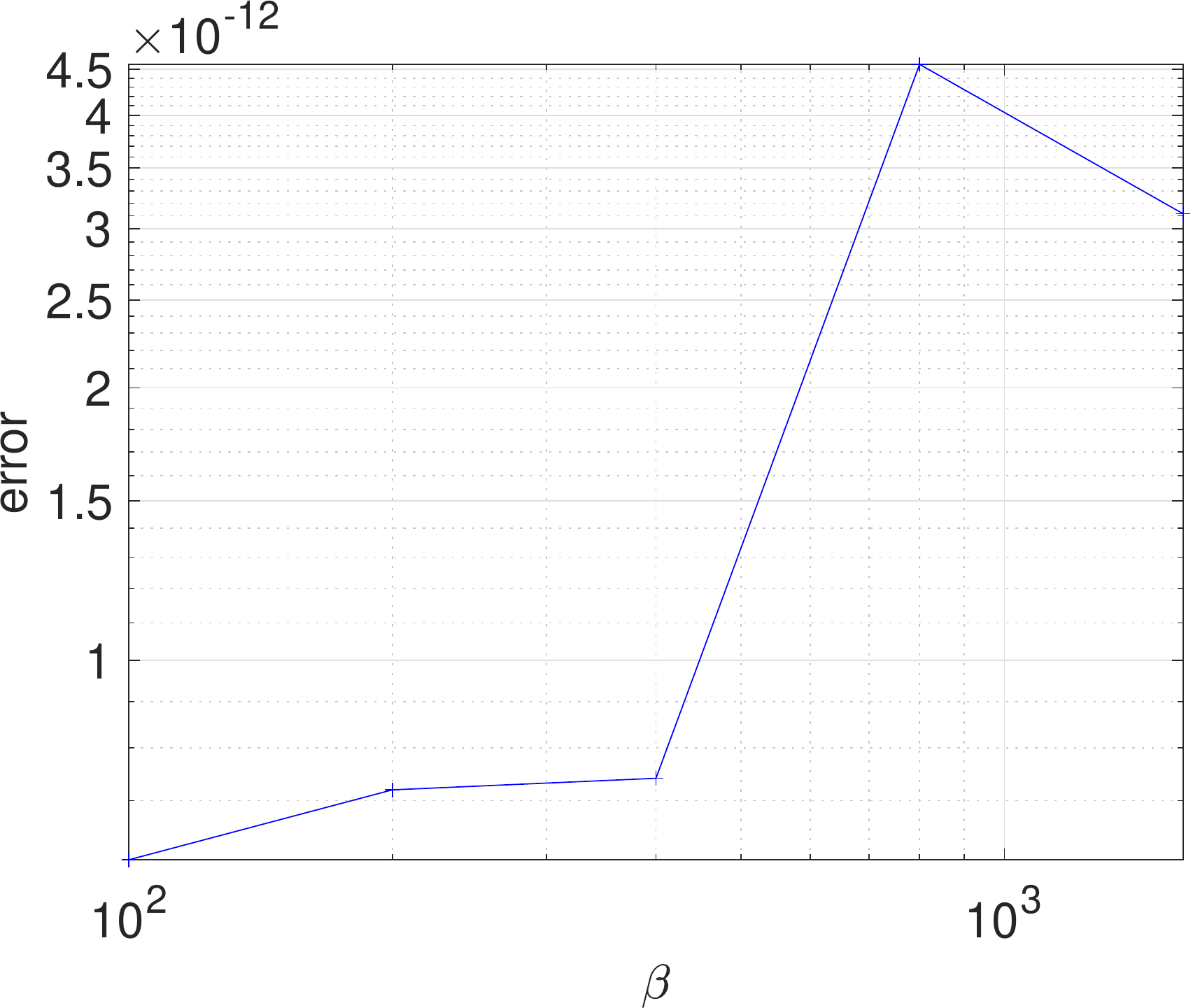}\\
    (c) & (d)
  \end{tabular}
  \caption{Fermi-Dirac operator. (a) $a(x)$ for $\beta=100$. (b) The degree $d$ of the polynomial
    approximation as a function of $\beta$.  (c) The total phase factor construction time in
    seconds as a function of $\beta$. (d) The relative $L_\infty$ norm error \eqref{eq:errest} as a
    function of $\beta$.}
  \label{fig:fermi}
\end{figure}

Finally, we consider the (shifted) Fermi-Dirac function
\[
f(x) = \frac{1 - e^{\beta x}}{1 + e^{\beta x}}.
\]
In the $t$ variable, it takes the form
\[
f(t) = \frac{1 - e^{\beta \cos(t)}}{1 + e^{\beta \cos(t)}}.
\]
The procedure mentioned above is used to compute a trigonometric approximation $a(t)$ to $f(t)$ (up
to a constant factor) with $\|a\|_\infty=0.3$. We perform the tests for $\beta =100, 200, 400,
800, 1600$ and the results are summarized in Figure \ref{fig:fermi}.

\section{Discussions}

In this paper, we proposed a new factorization algorithm for computing the phase factors of quantum
signal processing. The proposed algorithm avoids root finding of high degree polynomials by using a
key component of the Prony's method. The resulting algorithm is numerically stable in the double
precision arithmetics. We have demonstrated the numerical performance with several important
examples, including Hamiltonian simulation, eigenstate filtering, matrix inversion, and Fermi-Dirac
operator. For future work, the immediate question is to prove theoretically the stability of the
algorithm with the proposed choice of $b(t)$.

\section*{Acknowledgments:}
The author thanks Lin Lin for introducing the topic of quantum signal processing and the reviewers
for constructive comments.



\bibliographystyle{abbrvurl}
\bibliography{ref}

\end{document}